\documentstyle[aps,preprint]{revtex}

\begin{document}
\title{Static and Dynamical Anisotropy Effects in Mixed State of D - Wave
Superconductors}
\maketitle
\begin{center}{
$^*$D. Chang\footnote{chang@phys.nthu.edu.tw},
$^*$C.-Y. Mou\footnote{mou@phys.nthu.edu.tw},
$^\dagger$B. Rosenstein\footnote{baruch@phys.nthu.edu.tw},
and $^*$C. L. Wu\footnote{d833323@phys.nthu.edu.tw}

\vspace{12pt} 
$^*${Department of Physics,
National Tsing--Hua University,
Hsinchu, Taiwan, 30043, R.O.C.}

$^\dagger${Electrophysics Department,
National Chiao Tung University,
Hsinchu, Taiwan, 30043, R.O.C.}
}\end{center}
\date{The Date }

\begin{abstract}
We describe effects of anisotropy caused by the crystal lattice in d-wave
superconductors with s-wave mixing using the effective free energy approach.
Only the d-wave order parameter field $d$ is introduced, while the effect of
the s wave mixing as well as other effects breaking the rotational symmetry
down to the fourfold symmetry of the crystal are represented by a single
four (covariant) derivative term: $\eta d^{*}(\Pi _{y}^{2}-\Pi
_{x}^{2})^{2}d $. The single vortex solution in a phenomenologically
interesting range of parameters is almost identical to the two order
parameters approach. We analytically consider the most general oblique
lattice and orientation, but find that only rectangular body centered 
lattices are realized. A
critical value $\eta _{c} $ at which a phase transition from the rectangular 
lattice to the square lattice takes place. The influence on the phase
transition line is discussed. The formalism is extended to the time
dependent anisotropic Ginzburg-Landau equations in order to calculate the
effect of the anisotropy on the flux flow. The moving vortex structure is
established and the magnetization as function of the current is calculated.
Although the linear conductivity tensor is rotationally invariant due to the
fourfold symmetry, the nonlinear one shows anisotropy. We calculate
dependence of both direct and Hall I-V curves on the angle between the
current and the crystal lattice orientation.
\end{abstract}


\section{Introduction and Summary}

It is widely believed that the superconductivity in layered high $T_{c}$
cuprates is largely due to the $d_{(x^{2}-y^{2})}$ pairing \cite{Annett}.
The evidence for the d - wave pairing comes from variety of different
measurements. A partial list includes the $\mu $SR measurements of the
penetration depth \cite{Hardy}, quantum phase interference \cite{Tsuei},
angular resolved photoemission \cite{Shen}, thermal conductivity \cite{Aubin}%
, vortex lattice structure observed using neutron scattering \cite{Keimer}
and tunneling spectroscopy \cite{Maggio} and nuclear spin relaxation rate 
\cite{Martindale}. While most of these experiments directly probe the energy
gap and the low lying excitations, the vortex lattice observations are
different. They try to relate the structure and interactions of Abrikosov
vortices to the nature of the order parameter. This would be extremely
difficult if the order parameter would be a pure $d$ wave. The structure of
a single vortex and even the vortex lattice on the level of the
''macroscopic'' Ginzburg - Landau equations would be the same as for that of
the usual $s$ wave superconductors.

There are numerous indications, both theoretical \cite{Chubukov,Dagotto} and
experimental\cite{Tsuei,Martindale}, that even though the major pairing
mechanism is the $d$ wave pairing, there is a small admixture of the $s$
wave pairs in the condensate. Ren et al \cite{Ting1} using a
phenomenological microscopic model (in the weak-coupling limit) and Soininen
et al \cite{Berlinsky1} considering attractive nearest neighbors interaction
derived a two field effective Ginzburg-Landau (GL) type theory. The two
complex order parameter fields, $s$ and $d$ describe the gap function in
corresponding channels. The most general free energy of this kind reads \cite
{Berlinsky2,Joynt}:

\begin{eqnarray}
f &=&\alpha _{s}|s|^{2}-\alpha _{d}|d|^{2}+\beta _{1}|s|^{4}+\beta
_{2}|d|^{4}+\beta _{3}|s|^{2}|d|^{2}+\beta _{4}(s^{*2}d^{2}+d^{*2}s^{2}) 
\nonumber \\
&&+\gamma _{s}|{\bf \Pi }s|^{2}+\gamma _{d}|{\bf \Pi }d|^{2}+\gamma
_{v}[s^{*}(\Pi _{y}^{2}-\Pi _{x}^{2})d+c.c.]  \label{twofield}
\end{eqnarray}
where ${\bf \Pi }\equiv -i{\bf \nabla }-e^{*}{\bf A}$ is the covariant
derivative and \ $e^{*}$ is the charge of the Cooper pair (throughout this
paper we use the convention $c=\hbar =1$). \ Within a particular microscopic
model there might be some relations between these coefficients, but since
the ultimate microscopic theory is not known as yet, all of them should be
considered as phenomenologically fixed parameters.

Using equations following from this free energy or more fundamental
equations (see recent quasiclassical Eilenberger equations treatment in \cite
{Ichioka}), one obtains a characteristic four-lobe structure with four zeros
for the s - wave inside a single vortex \cite{Berlinsky2,Blatter}. Therefore
the vortex core looses the full rotational symmetry and only the fourfold $%
D_{4h}$ symmetry remains. The distribution of the magnetic field was also
obtained recently \cite{Blatter,Berlinsky2}. A somewhat different structure,
however, was found in another recent numerical work\cite{Ting3} (our
analytic calculation confirms that of \cite{Blatter,Berlinsky2} and
contradicts that of \cite{Ting3}).

Outside the core the s-wave vanishes, while the d-wave becomes rotationally
invariant, indistinguishable from the usual Abrikosov solution. Therefore,
to look for differences in the behavior of vortices one would like to be
closer to $H_{c2},$ so that the core will be more important (which is not
easy for high $T_{c}$ superconductors due to their large $H_{c2})$. However,
since the fourfold vortex core structure comes into a conflict with the high
symmetry of the triangular lattice, the asymmetry of vortices can distort
the usual triangular vortex lattice at already accessible fields much lower
than $H_{c2}$. Another phenomenon in which the vortex core plays a major
role is the dissipation in the course of the flux flow. These are the two
major phenomena in clean superconductors in which one can look for
anisotropy effects (we neglect pinning and other disorder and fluctuations,
and concentrate only on YBCO single crystals).

In this paper we study in detail the above two phenomena, vortex and the
vortex lattice structure and the flux flow, using a greatly simplified
model: time independent and time dependent one component effective Ginzburg
- Landau equations. This formulation, essentially without loss of
generality, allows us to avoid numerical methods and to extend and clarify
various delicate questions about the single vortex and the vortex lattice
structure for which there is some controversy or lack of concrete proof. In
particular the location of the phase transition to the square lattice is
calculated and depends just on one combination of the parameters. Moreover
the quantitative discussion can be extended to study moving flux lattices,
which, as is well known \cite{Dorsey,Hu}, are much more demanding, as far as
the calculational complexity is concerned. Their shape and orientation of
moving lattices (or large bundles) with respect to the crystal lattice and
electric field (or current) is determined. Then we calculate the current due
to the flux flow as function of the external electric field. Since the
deviations from the full rotational symmetry in the flux flow can be clearly
seen only in the nonlinear regime (the fourfold symmetry forces the full
rotational invariance of the Ohmic conductivity tensor) one has to go beyond
linear response.. The simplified formulation is indispensable in this case 
\cite{Dorsey,Hu}, but the result turns out to be remarkably simple. In the
rest of this section we outline the basic assumptions, methods and results
pointing out where in the paper more details can be found.

Within the two field formulation Soininen at al \cite{Berlinsky2} observed
that in predominantly d - wave superconductor the s - wave component is
generally very small: it is ''induced'' by variations of the larger d
component. Mathematically the dominance of the d - wave follows from the
fact that coefficient of the $d^{*}d$ term, $-\alpha _{d}=\alpha ^{\prime
}(T-T_{c})$, is negative , while that of the $s^{*}s$, $\alpha _{s}$, is
positive in the free energy, Eq.(\ref{twofield}). Therefore, it is the $d$
field which acquires a nonzero value. Then the rotationally noninvariant
derivative term $s^{*}(\Pi _{y}^{2}-\Pi _{x}^{2})d$ '' communicates'' the
deviations from the condensate value of $d$ to $s$. Note that this is the
only term up to (scaling) dimension three in the free energy which breaks
the full rotational symmetry. If its coefficient $\gamma _{v}$ is not very
large, the $s$ field never becomes comparable to $d$. Soininen et al \cite
{Berlinsky2} observed that even near the core, where $d$ is the smallest, it
is nevertheless larger then $s$ by a factor of 20 at least. This in
particular means that many small terms like $\left| s\right| ^{4}$are
irrelevant. The field $d$ is the critical field near $T^{c}$, while $s$ is
not and can be ''integrated out'' perturbatively. It will be explained in
some detail in Section II, that this generates an effective (scaling)
dimension five term for $d$, breaking the rotational symmetry, so that our
starting point to study the rotation symmetry breaking effects is an
effective free energy 
\begin{equation}
f_{eff}[d]=\frac{1}{2m_{d}}|\Pi d|^{2}-\alpha _{d}|d|^{2}+\beta |d|^{4}-\eta
d^{*}\left( \Pi _{y}^{2}-\Pi _{x}^{2}\right) ^{2}d.  \label{fe}
\end{equation}
Here we have replaced $\gamma _{d}$ by a more conventional notation $%
1/2m_{d} $ and parameter $\eta \equiv \gamma _{v}^{2}/\alpha _{s}$
quantifies the deviation from the rotational symmetry. Its relation to other
parameters in the two field approach is derived in Section II. We calculate
all the above mentioned rotational symmetry breaking effects to the first
order in $\eta $. This formulation follows the general philosophy of
effective free energy written in terms of critical fields only. The
noncritical fields just renormalize the coefficients. The rotational
symmetry breaking term has dimension five, but breaks the symmetry and is
therefore a ''dangerous irrelevant term'' using the terminology of critical
phenomena \cite{Amit}. The contributions to it might come not only from the
d - s mixing, which always give positive $\eta $, but also from other
sources. Even in conventional superconductors such effects exist \cite
{Takanaka}. In YBCO, twinning might be an important contribution to it. This
formulation avoids the problem of the second phase transition at $T_{s}$
that one encounters assuming $\alpha _{s}=\alpha ^{\prime }(T_{s}-T)$ in the
two field formalism$.$

The single vortex solution is obtained in Section III. It is almost
identical to the solutions obtained earlier within the framework of the two
order parameters theory, One still can define the ''effective s-wave field
by $s=-\frac{\gamma _{v}}{\alpha _{s}}(\Pi _{y}^{2}-\Pi _{x}^{2})d$ and
observe the four lobe structure, see Fig. 2 and 3 for $d$ and $s$ components
respectively. Relation to earlier work (discrepancies or common points) are
summarized in the Appendix.

The vortex lattice near $H_{c2}$ is studied comprehensively in Section IV.
The simplicity of the formulation allows for an analytic study of all the
possibilities, not considered before or considered using uncontrollable
approximations. The degrees of freedom we include in the analysis contain:
(1) an arbitrary rotation angle $\varphi $ between the crystal lattice and
the vortex lattice (Fig. 4), \ and (2) all the possible lattice, not only
the rectangular ones considered before 
(\cite{Berlinsky1},\cite{Berlinsky2},\cite
{Ting3}). Instead of using the variational method to solve the linearized
set of equations, we solve it exactly even in the moving lattice case.\ This
is the first time that the lattice is\ demonstrated to be rectangular 
body centered using the
most general lattice in the analysis. We tabulate the lattice
characteristics for different $\eta $ in Fig. 7. At certain value of $\eta $
there is a phase transition from rectangular to a more symmetric square lattice
first noticed \cite{Berlinsky2}. The exience of a phase transion becomes 
obvious in our formulation in which the effective strength of
the four fold symmetry is proportional to the magnetic field,
characterized by a dimensionless parameter
$\eta^{\prime}\equiv \eta m_{d}e^{*}H$.
In low fields, the four fold symmetry is subdominant, so
the lattice is closer to the triangle lattice. In high fields,
the four fold symmetry dominates, so the lattice becomes square. 
We find that the transition occurs at
$\eta_{c}^{\prime}=.0235.$
The upper critical field line $H_{c2}(T)$ is given in
Eq.(\ref{hc2t}). However at the end of this Subsection IV.C we caution
against too direct interpretation of this result by considering other
possible effects which are isotropic and also contribute to the curvature.

The moving lattice solutions are derived in Section V from a time dependent
generalization of Eq.(\ref{fe}). They are not only needed for the nonlinear
conductivity calculation which follows, but are also interesting in their
own right, since they are, in principle, observable. In the one field
formalism there is only one additional parameter to be added to take the
dissipation effects into account: the coefficient of the time derivative
term of $d$. Unlike in the case of the pure s-wave superconductor, the
moving (with arbitrary, not infinitesimal, velocity) lattice solution in
this case can not be obtained from the static one by a simple Galilean boost 
\cite{Hu}. It is a nontrivial problem and we were able to solve it
perturbatively in $\eta $. Unlike the s - wave moving lattices (which are
triangular \cite{Hu}), orientation is determined by the direction of the
crystal lattice as well as by the current direction. The ''dynamical phase
transition line as function of current and its orientation with respect to
the atomic lattice are quantitatively discussed in Subsection V.B and the
result is given in Eqs.(\ref{line},\ref{line1},\ref{line2}). Magnetization
close to $H_{c2}$ (or the Abrikosov $\beta _{A})$ is a very simple function
of the current $J$ and its direction. The result for the Abrikosov $\beta
_{A}$ is: 
\begin{equation}
\beta _{A}({\bf E})=\beta _{A}^{0}+c\eta E^{2}|\cos 2\Theta |,
\label{magnetization}
\end{equation}
where $\Theta $ is an angle between the electric field ${\bf E}$ and an axis
of the underlying atomic lattice axis. The number $\beta _{A}^{0}$
(typically a bit larger then 1) is the usual Abrikosov $\beta $ parameter
for a given lattice defined in Eq.(\ref{beta0}) and the constant $c$ is
given in Subsection V.C (Eqs.(\ref{beta1},\ref{G})).

Corrections to the linear conductivity tensor (which cannot break the
rotational symmetry) are briefly discussed and a detailed calculation of the
effect of the anisotropy on the nonlinear flux flow are given in Section VI.
The result is remarkably simple. In addition to isotropic linear part there
is an anisotropic cubic in $E$ term is:

\begin{equation}
\Delta {\bf J}=\eta \frac{2m_{d}\gamma ^{3}E^{3}}{\beta _{A}^{0}e^{*}H^{4}}%
\left( 1+\cos 4\Theta \right)  \label{Jdir}
\end{equation}
and the Hall current is

\begin{equation}
\Delta {\bf J}_{Hall}=-\eta \frac{2m_{d}\gamma ^{3}E^{3}}{\beta
_{A}^{0}e^{*}H^{4}}\sin 4\Theta  \label{JHall}
\end{equation}
The two nonlinear contributions to currents are simply related. In these
expressions $\gamma $ is the coefficient of the time derivative term in time
dependent Ginzburg - Landau equation Eq.(\ref{TDGL}). Both direct and Hall
I-V curves depend on the angle between the current and the crystal lattice
orientation via the fourth harmonic only. The result, contains only cubic
dependance of the currents on the electric field, higher orders being
cancelled.

Finally in Section VII we conclude by briefly discussing possible
experiments to observe various above mentioned effects, as well as some
generalizations.

\section{Time independent and time dependent effective Ginzburg - Landau
equations.}

On very general grounds superconductivity is sometimes considered as a
phenomenon of ''spontaneous $U(1)$ gauge symmetry breaking '' \cite{Anderson}
irrespective of the mechanism of pairing or channels in which it occurs. The 
$U(1)$ electric charge symmetry should be represented by a single order
parameter: the superconducting $U(1)$ phase. Mechanisms of
superconductivity, as far as connection to the gap functions appearing in
the microscopic description is concerned, may differ, but this general order
parameter representation of the superconducting phase remains the same.
While in pure s - wave or d - wave superconductors the $U(1)$ phase is
simply identified with the phase of the gap function, in more complicated
microscopic theories with few channels opened, the superconducting phase is
just the common phase of various gap functions. Therefore, quantities other
than the phase which enter various phenomenological Ginzburg-Landau (GL)
type equations, although useful, are not directly related to the spontaneous
gauge symmetry breaking.

The s-d mixing two component GL free energy Eq.(\ref{fe}) leads to the
following set of equations: 
\begin{eqnarray}
\left( \gamma _{d}\Pi ^{2}-\alpha _{d}\right) d+\gamma _{v}(\Pi _{y}^{2}-\Pi
_{x}^{2})s+2\beta _{2}|d|^{2}d+\beta _{3}|s|^{2}d+2\beta _{4}s^{2}d^{*} &=&0
\label{d-equation} \\
(\gamma _{s}\Pi ^{2}+\alpha _{s})s+\gamma _{v}(\Pi _{y}^{2}-\Pi
_{x}^{2})d+2\beta _{1}|s|^{2}s+\beta _{3}|d|^{2}s+2\beta _{4}d^{2}s^{*} &=&0
\label{s-equation}
\end{eqnarray}

As we discussed in the Introduction, the noncritical $s$ component is
induced by the $d-s$ mixing gradient term. Therefore the length scale of the
variations of the field $s$ is the d = wave's coherence length $\xi _{d}=%
\sqrt{\frac{\gamma _{d}}{\alpha _{d}}}$. Consequently the derivative term in
Eq.(\ref{s-equation}) $\gamma _{s}\Pi ^{2}s\sim \left( \gamma _{s}/\xi
_{d}^{2}\right) s$ is small compared to $\alpha _{s}s$. This requirement $%
\frac{\gamma _{s}}{\gamma _{d}}\frac{\alpha _{d}}{\alpha _{s}}\ll 1$
(typically $\frac{\gamma _{s}}{\gamma _{d}}\sim 1$) holds for the vortex
solution of \cite{Berlinsky2,Ting2,Blatter} and is, in fact, an excellent
approximation in both near and far from the core regions. Then, according to
Eq.(\ref{s-equation}), the field $s$ is, to the first order in $1/\alpha
_{s} $:

\begin{equation}
s\approx -{{\frac{\gamma _{v}}{\alpha _{s}}}}(\Pi _{y}^{2}-\Pi _{x}^{2})d
\end{equation}
Substituting this equation back to Eq.(\ref{d-equation}), we obtain, to
first order in $1/\alpha _{s}$, 
\begin{equation}
\left( \frac{1}{2m_{d}}\Pi ^{2}-\alpha _{d}\right) d-\eta (\Pi _{y}^{2}-\Pi
_{x}^{2})^{2}d+2\beta |d|^{2}d=0  \label{GL}
\end{equation}
where $\gamma _{d}$ was replaced by a more standard notation $\frac{1}{2m_{d}%
}$ and $\eta \equiv {{\frac{\gamma _{v}^{2}}{\alpha _{s}}}}$. The second
term should be treated as a perturbation. In principle there are terms of
higher order in $1/\alpha _{s}$, but one cannot take them consistently into
account without simultaneously including additional terms in the original
two field GL equations. From Eq.(\ref{GL}) the effective free energy Eq.(\ref
{fe}) follows. Of course the effective free energy $f_{eff}$ Eq.(\ref{GL})
is valid only if possible higher orders of the field $d$ and derivatives are
neglected.

Note that even the linearized set of Eq.(\ref{d-equation},\ref{s-equation})
is highly nontrivial. Authors of Ref. \cite{Berlinsky2} resort to the
variational estimate to find a solution. On the other hand, the linearized
Eq.(\ref{GL}) can be easily solved perturbatively in $1/\alpha _{s}$.
Another advantage of this equation, especially as far as relation to
experiments is concerned, is that the number of coefficient is much smaller.
Instead of 6 additional parameters in the two field free energy, there is
just one additional adjustable parameter compared to the usual s - wave GL
equations.

The effective free energy approach can be also motivated by considering the
fluctuations of the two fields theory. Assuming that only the $d$ field is
critical, one can integrate out the $s$ field fluctuations perturbatively.
The lower order terms coming out from this analysis are precisely the same
as $f_{eff}$ as expected.\ In principle, the coefficients of the effective
one field free energy should be fixed by a microscopic theory in the same
spirit as the way that the coefficients of the two component equations
should be fixed. The general form of the effective free energy can be
obtained just by the dimensional analysis and symmetry. Generally, we should
allow all the terms invariant under the group $D_{4h}$ with dimensions five
or less. It is a well known property of the $D_{4h}$ symmetry that at the
level of the dimension three (relevant) terms, full rotational symmetry is
restored.

Therefore, one has to consider ''irrelevant'' (scaling) dimension five
gradient terms in order to break the rotational symmetry down to $D_{4h}$. 
{\it Apriori} there are five such terms: $d^{*}\Pi _{x}^{4}d$, $d^{*}\Pi
_{y}^{4}d$, $d^{*}\Pi _{x}^{2}\Pi _{y}^{2}d$, $d^{*}\Pi _{x}\Pi _{y}^{3}d$,
and $d^{*}\Pi _{x}^{3}\Pi _{y}d$. The last two break the reflection
symmetry, while out of the remaining three, one can only construct two
linear combinations invariant under the rotation of $\pi /2$. They are $%
d^{*}(\Pi _{y}^{2}-\Pi _{x}^{2})^{2}d$ and $d^{*}(\Pi ^{2})^{2}d$. The
second term is fully rotational invariant, and therefore is not important
for studying anisotropy effects. It, however, contributes to effects not
directly related to anisotropy such as the shape of the phase transition
line. We will come back to this point in Section IV. Similarly, all the
other dimension five terms that one should in principle include are
rotationally invariant. They just add a small contribution to the
rotationally invariant parts of physical quantities and need not be included
when studying the rotational symmetry breaking effects.

The time dependent GL equation in the one field formulation is also quite
simple: 
\begin{equation}
\gamma \left( \frac{\partial }{\partial t}+ie^{*}\Phi \right) d=-\left( 
\frac{1}{2m_{d}}\Pi ^{2}-\alpha _{d}\right) d+\eta (\Pi _{y}^{2}-\Pi
_{x}^{2})^{2}d-2\beta |d|^{2}d,  \label{TDGL}
\end{equation}
where $\Phi $ is the electric potential. The above equation describes the
time evolution of the order parameter and will be used to describe moving
vortex systems. It involves only one additional parameter $\gamma $ compared
to the 2$\times $2 matrix for the two field formalism. Although, in
principle, this parameter describing various dissipation effects can have a
complex part \cite{Dorsey}, we will consider only real values.

\section{The single vortex solution}

In this section we find an isolated vortex solution of the one component
equation Eq.(\ref{GL}) near $H_{c1}$.The opposite case of magnetic field
close to $H_{c2}$ will be considered in next section. We measure the field
in units of the vacuum expectation value $\Psi _{0}=\sqrt{\alpha _{d}/2\beta
_{2}},$ and length in units of the coherence length $\xi _{d}=1/\sqrt{%
2m_{d}\alpha _{d}}$ $.$ In strongly type II materials (when the
Ginzburg-Landau parameter $\kappa $ is very large), as is the case in high $%
T_{c}$ superconductors, we can safely ignore the magnetic field and the
dimensionless GL\ equation becomes: 
\[
(-\nabla ^{2}-1)d-\lambda (\nabla _{y}^{2}-\nabla _{x}^{2})^{2}d+|d|^{2}d=0 
\]
where $\lambda \equiv 4\eta \,m_{d}^{2}\alpha _{d}$ is our dimensionless
small perturbative parameter.. We solve it perturbatively in $\lambda $ as
follows. Let $d=d_{0}+\lambda d_{1}$ , where $d_{0}=f_{0}(r)e^{i\phi }$ is
the solution of the standard unperturbed GL equation. Then the first order
equation in $\lambda $ is 
\begin{equation}
(-\nabla ^{2}-1)d_{1}+(2|d_{0}|^{2}d_{1}+d_{0}^{2}d_{1}^{*})=(\nabla
_{y}^{2}-\nabla _{x}^{2})^{2}d_{0}  \label{effective equation 1st order}
\end{equation}
The angular dependence of $d_{1}$ is easily observed to contain only three
harmonics: $e^{-3i\phi }$, $e^{+i\phi }$ and $e^{5i\phi }$. This is
consistent with the fourfold symmetry which is built into the theory. We
therefore decompose $d_{1}$ into combination of these three harmonics: 
\begin{equation}
d_{1}(r,\phi )=f_{-3}(r)e^{-3i\phi }+f_{1}(r)e^{i\phi }+f_{5}(r)e^{5i\phi }
\end{equation}

The equation becomes 
\begin{eqnarray}
\left( 
{\displaystyle {d \over dr^{2}}}%
^{2}+%
{\displaystyle {1 \over r}}%
{\displaystyle {d \over dr}}%
-%
{\displaystyle {9 \over r^{2}}}%
+1\right) f_{-3}-f_{0}^{2}(2f_{-3}+f_{5}) &=&-J_{-3}(r)  \label{f3} \\
\left( 
{\displaystyle {d \over dr^{2}}}%
^{2}+%
{\displaystyle {1 \over r}}%
{\displaystyle {d \over dr}}%
-%
{\displaystyle {25 \over r^{2}}}%
+1\right) f_{5}-f_{0}^{2}(2f_{5}+f_{-3}) &=&-J_{5}(r)  \label{f5} \\
\left( 
{\displaystyle {d \over dr^{2}}}%
^{2}+%
{\displaystyle {1 \over r}}%
{\displaystyle {d \over dr}}%
-%
{\displaystyle {1 \over r^{2}}}%
+1\right) f_{1}-3f_{0}^{2}f_{1} &=&-J_{1}(r)  \label{f1}
\end{eqnarray}
with $J_{i}$ defined by:

\[
(\nabla _{y}^{2}-\nabla _{x}^{2})^{2}\left[ f_{0}(r)e^{i\phi }\right]
=e^{i\phi }J_{1}(r)+e^{-3i\phi }J_{-3}(r)+e^{5i\phi }J_{5}(r) 
\]
As is well known, the analytic expression for $f_{0}$ does not exist, \
however, there are a number of known good approximations. Using one of them%
\cite{Tinkham}, $f_{0}(r)=\frac{r}{\sqrt{r^{2}+\xi _{v}^{2}}}$ , the set of
linear equations are then solved numerically (the third equation decouples
from the first two). The results are shown in Fig.1. The d - wave
configuration is basically indistinguishable from that of the two fields
formalism for $\lambda =0.15$, see Fig. 2.

Note also that within the same approximation and normalization, the $s$
component is: 
\begin{equation}
s=\lambda ^{\prime }(\nabla _{y}^{2}-\nabla _{x}^{2})d_{0}
\end{equation}
where $\lambda ^{\prime }=2\gamma _{v}m_{d}(\alpha _{d}/\alpha _{s})=\frac{%
\lambda }{2\gamma _{v}m_{d}}$ is another dimensionless small parameter. It's
easy to see that $s$ has the asymptotic behavior 
\begin{equation}
\begin{tabular}{ll}
$s\sim re^{-i\phi }$ & $r\rightarrow 0$ \\ 
$s\sim 
{\displaystyle {1 \over r^{2}}}%
e^{+3i\phi }$ & $r\rightarrow \infty $%
\end{tabular}
\label{s component}
\end{equation}

The $s$ field is plotted in Fig.3. The different winding numbers in near and
far asymptotic regions give rise to four additional poles in the s component
in the intermediate region. This confirms calculations of \cite{Berlinsky2}
even though some asymptotic analytic expression used there to obtain the
numerical results disagree with ours. The comparison with \ \cite{Berlinsky2}
and \cite{Ting2}\ is presented in Appendix.

\section{The Vortex Lattice near $H_{c2}$}

In this Section we follow a generalization of the Abrikosov's procedure \cite
{Abrikosov,SaintJames} to investigate the structure of the vortex lattice
near $H_{c2}$. One first ignores the non-linear terms in the GL equation and
finds a set of the lowest energy solutions $\Psi _{k_{n}}(x,y)$of the
linearized equation. The lattice solution is constructed as a linear
superposition 
\begin{equation}
d(x,y)=\sum_{n}C_{n}\Psi _{k_{n}}(x,y)
\end{equation}
in such a way that it is invariant under the corresponding symmetry group of
the lattice. It is well known that the free energy near $H_{c2}$ is
monotonic in the Abrikosov's parameter $\beta _{A}$, which is defined by $%
\beta _{A}=\langle |d|^{4}\rangle /\langle |d|^{2}\rangle ^{2}$. In the last
step needed for some applications, the overall normalization of the order
parameter is variationally fixed by minimizing the free energy including
non-linear terms.

A general lattice in 2D can be specified by three parameters $a,b$ and $%
\alpha $, where $a$ and $b$ are the two lattice constants, while $\alpha $
is the angle between the two primitive lattice vectors (Fig 4). Flux
quantization constrains them, so that $Hab\sin \alpha =\Phi _{0}$. In the
d-wave superconductors the rotational symmetry is broken, therefore the
relative orientation of the vortex lattice to the underlying lattice becomes
important. Later we will denote $\varphi $ to be the angle between $%
\stackrel{\rightarrow }{a}$ and one of the axes of the underlying lattice.
In Abrikosov's original paper \cite{Abrikosov} he assumed $C_{n}=C_{n+1}$
and obtained the square lattice, later Kleiner {\it et al }generalized the
procedure to the case where $C_{n}=C_{n+2}$. In this way all the rectangular
body centered
lattices can be included in the analysis. In previous works on d-wave
superconductivity ref. .\cite{Berlinsky2,Ting2}, the same formalism was
used, however, this does not include the most general lattice. \ In this
section we follow a more generalized formulation of Ref. \cite{SaintJames}
which covers all possible lattice types.

In our case we first solve the linearized GL equation pertubatively in the
anisotropy parameter $\eta $. And then we obtain an analytic expression of $%
\beta _{A}$ as a function of $a$, $b$ (or $\alpha $) and $\varphi $ .
Finally, we minimize the free energy analytically with respect to $\varphi $
and numerically with respect to $a$, $b$ (or $\alpha $) to find the lattice
structure.

\subsection{The perturbative solution of the linearized GL equations}

We start from the effective linearized GL equation Eq.(\ref{GL}) 
\begin{equation}
{\displaystyle {1 \over 2m_{d}}}%
\Pi ^{2}d-\eta \,(\Pi _{y}^{2}-\Pi _{x}^{2})^{2}d=\alpha _{d}d,
\label{linearGL}
\end{equation}
where for later convenience we have moved $\alpha _{d}d$ to the right hand
side. It is important to note that in Eq.(\ref{linearGL}) we have assumed
that the coordinate system and the underlying microscopic lattice coincide.
Later it will be convenient to orient the coordinate system $(x,y)$ with the
Abrikosov vortex lattice rather the atomic crystal. In general, if the
crystal is rotated by an angle $\varphi $ counterclockwise with respect to
the coordinate system, Eq.(\ref{linearGL}) becomes 
\begin{equation}
{\displaystyle {1 \over 2m_{d}}}%
\Pi ^{2}d-\eta \,\left[ \cos 2\varphi \,(\Pi _{x}^{2}-\Pi _{y}^{2})+\sin
2\varphi \,(\Pi _{x}\Pi _{y}+\Pi _{y}\Pi _{x})\right] ^{2}d=\alpha _{d}d.
\label{linearGL1}
\end{equation}
It is convenient to introduce dimensionless creation and annihilation
operators defined by 
\begin{eqnarray*}
\hat{a} &=&\frac{i\Pi _{+}}{\sqrt{2}}l_{H}, \\
\hat{a}^{\dagger } &=&\frac{-i\Pi _{-}}{\sqrt{2}}l_{H},
\end{eqnarray*}
where $\Pi _{\pm }\equiv \Pi _{x}\pm i\Pi _{y}$ and the scaling parameter $%
l_{H}=1/\sqrt{e^{*}H}$ is the magnetic length. In terms of $\hat{a}$ and $%
\hat{a}^{\dagger }$, Eq.(\ref{linearGL}) becomes 
\begin{equation}
\left[ \hat{a}^{\dagger }\hat{a}+\frac{1}{2}-\eta ^{\prime }\,(e^{+2i\varphi
}\hat{a}^{\dagger 2}+e^{-2i\varphi }\hat{a}^{2})^{2}\right] d(x,y)=\frac{%
H^{0}}{2H}d(x,y).  \label{GLina}
\end{equation}
Here dimensionless parameter $\eta ^{\prime }$ is given by $\eta ^{\prime
}=\eta m_{d}e^{*}H.\ $For later convenience, we have defined an unperturbed
(conventional) upper critical fields $H^{0}\equiv \Phi _{0}/(2\pi \xi
_{d}^{2})=2m_{d}\alpha _{d}/e^{*}$.

It is convenient to choose the Landau gauge ${\bf A=}Hx\widehat{y}$, since
in this gauge the momentum operator ${\rm \hat{p}}_{y}$ commutes with $\hat{a%
}$. Therefore we can choose $d(x,y)$ to be an eigenvector of ${\rm \hat{p}}%
_{y}$ with eigenvalue $k$: $d(x,y)=\exp (iky)\psi _{k}(x)$. The operators $%
\hat{a}$ and $\hat{a}^{\dagger }$ in this gauge become 
\begin{eqnarray}
\hat{a} &=&\frac{1}{\sqrt{2}}\left( \frac{d}{d\tilde{x}}+\tilde{x}\right) ,
\label{a} \\
\hat{a}^{\dagger } &=&\frac{1}{\sqrt{2}}\left( -\frac{d}{d\tilde{x}}+\tilde{x%
}\right) ,  \label{a2}
\end{eqnarray}
where $\tilde{x}\equiv (x-x_{0})/l_{H}$ is dimensionless \ with $x_{0}\equiv
kl_{H}^{2}$. Standard perturbation theory for the Schr\"{o}dinger type
equations gives for the lowest\ eigenvalue: 
\begin{equation}
\frac{H^{0}}{2H}=\frac{1}{2}-2\eta ^{\prime }+O(\eta ^{2}),  \label{result1}
\end{equation}
This determines the upper critical field. Note that the relative angle $%
\varphi $ does not affect $H$ in the lowest order. We will come back to
examine this result more closely later. The corresponding eigenfunction $%
\psi (x)$ is: 
\begin{eqnarray}
\psi (\tilde{x}) &\equiv &\psi _{0}+\eta ^{\prime }\psi _{1}=|0\rangle +\eta
^{\prime }\frac{\,e^{4i\varphi }}{4}\sqrt{4!}\,|4\rangle +O(\eta ^{2})
\label{perturbedd} \\
&=&\left( \frac{1}{\pi l_{H}^{2}}\right) ^{1/4}\left[ 1+\eta ^{\prime }\frac{%
e^{+4i\varphi }}{16}H_{4}(\tilde{x})\right] \exp (-\frac{\tilde{x}^{2}}{2}).
\end{eqnarray}
where $H_{4}(x)$ is the forth Hermit polynomial.

\subsection{The slope of the upper critical field in d-wave superconductors}

From Eq.(\ref{result1}) we obtain that the upper critical field $H$
satisfies 
\[
H^{0}=H-4\eta m_{d}e^{*}H^{2}. 
\]
Solving it perturbatively, we get

\[
H=H^{0}+4\eta m_{d}e^{*}(H^{0})^{2}. 
\]
Reinstating the temperature dependence of coefficients of the GL equation,
in the simplest case $\alpha (T)=\alpha ^{\prime }(T_{c}-T)$, we finally get

\begin{equation}
H(T)=\frac{2m_{d}\alpha ^{\prime }}{e^{*}}\left[ \left( T_{c}-T\right)
+8\eta m_{d}^{2}\alpha ^{\prime }\left( T_{c}-T\right) ^{2}\right] .
\label{hc2t}
\end{equation}

We observe that around $T_{c}$ for a positive $\eta $ the $H(T)$ curve bends
upwards, in agreement with the two field results \cite{Joynt,Berlinsky2}.
This effect has been reported in some experiments. However one should be
cautioned against taking this too seriously. First, as we will discuss in
some detail later, the coefficient $\eta $ should not necessarily be
positive in all the samples. For example, twinning is expected to give a
negative contribution to it. For negative $\eta $ correction to the
curvature changes sign. Second, although in this study we concentrate on the
effects of the anisotropy, which is represented by the (scaling) dimension
five four derivative term $d^{*}(\Pi _{y}^{2}-\Pi _{x}^{2})^{2}d,$ as we
discussed in Section II, there are other rotationally invariant terms of the
same dimensionality. The second dimension five four derivative term, $\tau $$%
d^{*}(\Pi ^{2})^{2}d,$ gives contribution similar to Eq.(\ref{hc2t}). One
does a calculation similar to that we performed for $d^{*}(\Pi _{y}^{2}-\Pi
_{x}^{2})^{2}d$ and obtains:

\[
H_{c2}(T)=\frac{2m_{d}\alpha ^{\prime }}{e^{*}}\left[ \left( T_{c}-T\right)
+8m_{d}^{2}\alpha ^{\prime }(\eta -\tau )\left( T_{c}-T\right) ^{2}\right] . 
\]

This means that for positive $\tau $ the sign of the correction might be
changed. Unlike $\eta $ which can be fixed by rotation symmetry breaking
effects experiments, this rotationally invariant correction is more
difficult to estimate phenomenologically.

\subsection{The Abrikosov parameter calculation}

Now we proceed to calculate the Abrikosov's $\beta _{A}\equiv
<|d|^{4}>/<|d|^{2}>^{2}$. Here the angular bracket is defined as the average
over the 2D volume, i.e., $<f>\equiv \frac{1}{V}\int d^{2}{\bf r}f({\bf r})$%
, and $V$ is the total volume of the system. If the function $f({\bf r})$ is
periodic , it is sufficient to calculate the average over one unit cell.

In the gauge we have chosen in the previous section, a generic solution of
the linearized equation takes the form $\Psi _{k}(x,y)=\exp (iky)\,\psi
(x-kl_{H}^{2})$. Periodicity in $y$-direction (our lattice vector ${\bf a}$
by assumption is aligned with this axis, see Fig.4) allows the following
linear combinations: 
\begin{equation}
d(x,y)=\sum_{n=-\infty }^{n=\infty }C_{n}\Psi _{k}(x,y)=\sum_{n=-\infty
}^{n=\infty }C_{n}\exp \left( i\frac{2\pi n}{a}y\right) \psi \left( x-n\frac{%
2\pi l_{H}^{2}}{a}\right) ,  \label{linearcombination}
\end{equation}
If the second lattice constant is $b$ and it makes an angle $\alpha $
relative to the y axis, the periodicity in $\hat{b}$ direction requires that 
$d(x-b\sin \alpha ,y+b\cos \alpha )=d(x,y)$ (up to a phase). One can achieve
it by setting $b\sin \alpha =p\Delta x=p\left( 2\pi l_{H}^{2}/a\right) $ and 
$C_{n+p}=C_{n}\exp (i2\pi nb\cos \alpha /a)$, where $p$ is an integer. For
simple Bravais lattices, there is only one vortex in each unit cell.
Therefore, we can take $p=1$. The area of the unit cell is $ab\sin \alpha
=\Phi _{0}/H=2\pi l_{H}^{2}.$

As a result, all $C_{n}$ can be fixed up to an overall constant, to be fixed
later: 
\begin{equation}
C_{n}=\exp \left[ 2\pi i\frac{b}{a}\cos \alpha \frac{n(n-1)}{2}\right]
\label{C_n}
\end{equation}
and the wave function becomes: 
\begin{equation}
d(x,y)=\sum_{n=-\infty }^{n=\infty }\exp \left[ 2\pi i\frac{b}{a}\cos \alpha 
\frac{n(n-1)}{2}\right] \exp \left[ i\frac{2\pi n}{a}y\right] \psi (x-nb\sin
\alpha ),
\end{equation}
It is convenient to use new rectilinear coordinates whose axes coincide with
the vortex lattice directions (Fig.4). We shall denote them as $X$ and $Y$.
Their relations to the old $x-y$ coordinates are $y=Y+X\cos \alpha $ and $%
x=-X\sin \alpha $.

The average of $|d|^{2}$ is then found by integrating $|d|^{2}$ over $0<X<b$%
\ and\ $0<Y<a$. The integration over $Y$ enforces a delta function and
simplifies the double summation to 
\begin{equation}
\langle |d|^{2}\rangle =\frac{1}{ab\sin \alpha }\sum_{n=-\infty }^{\infty
}a\sin \alpha \int_{0}^{b}dX|\psi [(-X-nb)\sin \alpha ]|^{2}.
\end{equation}
The summation over $n$ converts the integration domain into $(-\infty
,\infty )$. We thus obtain 
\begin{equation}
\langle |d|^{2}\rangle =\frac{|C_{0}|^{2}}{b\sin \alpha }\int_{-\infty
}^{\infty }dx\,|\psi (x)|^{2}.  \label{qudraticd}
\end{equation}
A similar manipulation on $\langle |d|^{4}\rangle $ leads to 
\begin{eqnarray}
&\langle |d|^{4}\rangle =&\frac{1}{ab\sin \alpha }\sum_{n,m,n^{\prime
},m^{\prime }}a\sin \alpha \delta _{m+m^{\prime },n+n^{\prime }}\,\exp
\left[ 2\pi i\frac{b}{a}\cos \alpha \frac{-m^{2}-m^{\prime
2}+n^{2}+n^{\prime 2}}{2}\right]  \nonumber \\
&&\times \int_{0}^{b}dX\psi ^{*}[(-X-mb)\sin \alpha ]\,\psi
^{*}[(-X-m^{\prime }b)\sin \alpha ]\psi [(-X-nb)\sin \alpha ]\,\psi
[(-X-n^{\prime }b)\sin \alpha ]\,  \nonumber \\
&=&\frac{1}{b\sin \alpha }\sum_{n,m,n^{\prime },m^{\prime }}\delta
_{m+m^{\prime },n+n^{\prime }}\,\exp \left[ 2\pi i\frac{b}{a}\cos \alpha 
\frac{-m^{2}-m^{\prime 2}+n^{2}+n^{\prime 2}}{2}\right]  \nonumber \\
&&\times \int_{-b\sin \alpha }^{0}dx\psi ^{*}(x-mb\sin \alpha )\,\psi
^{*}(x-m^{\prime }b\sin \alpha )\,\psi (x-nb\sin \alpha )\,\,\psi
(x-n^{\prime }b\sin \alpha )\,\,.  \label{d4}
\end{eqnarray}

Eqs.(\ref{qudraticd}) and (\ref{d4}) are general expressions for $\langle
|d|^{2}\rangle $ and $\langle |d|^{4}\rangle $. We now specialize them to
our perturbed $d$ field solution Eq.(\ref{perturbedd}). It is easy to see
that the correction to $\langle |d|^{2}\rangle $ starts from $\eta ^{\prime
2}$. We found that 
\begin{equation}
\langle |d|^{2}\rangle =\frac{1}{b\sin \alpha }\left( 1+\frac{3}{2}\eta
^{\prime 2}\right) .  \label{d2}
\end{equation}
The first order term vanishes, because according to Eq.(\ref{qudraticd}), it
is proportional to the inner product of $\psi _{0}$ and $\psi _{1\text{.}}$
Since we shall be interested only in $O(\eta ^{\prime })$ corrections, we
will drop this second order term.

The calculation of $\langle |d|^{4}\rangle $ is more involved even in zeroth
order \cite{SaintJames}. Because of the presence of the Kronecker delta,
there are only three independent summations in Eq.(\ref{d4}). We choose the
summation variables to be 
\begin{eqnarray}
Z &=&n+n^{\prime }=m+m^{\prime }  \nonumber \\
N &=&n-n^{\prime }  \nonumber \\
M &=&m-m^{\prime }  \label{summation variable}
\end{eqnarray}
Note that the new discrete variables $Z,M$ and $N$ are not completely
independent since they have to be either all even or all odd simultaneously.
The summation in Eq.(\ref{d4}) then becomes 
\begin{equation}
\sum_{m,m^{\prime },n,n^{\prime }}\delta _{m+m^{\prime },n+n^{\prime
}}=\sum_{\text{even }Z}\sum_{\text{even }M}\sum_{\text{even }N}+\sum_{\text{%
odd }Z}\sum_{\text{odd }M}\sum_{\text{odd }N}.  \label{summation}
\end{equation}
To zeroth order in $\eta $, the integrand in Eq.(\ref{d4}), after
appropriate rearrangement, has a simple Gaussian form 
\begin{equation}
\exp \left[ -\frac{2\sin ^{2}\alpha }{l_{H}^{2}}\left( X+\frac{Z}{2}b\right)
^{2}\right] \exp \left\{ -\frac{b^{2}\sin ^{2}\alpha }{l_{H}^{2}}\left[
\left( \frac{M}{2}\right) ^{2}+\left( \frac{N}{2}\right) ^{2}\right]
\right\} .
\end{equation}
As before, the summation over $Z$ in Eq.(\ref{summation}) extends the range
of the integral over $X$ to $(-\infty ,\infty ),$ so that the gaussian
integral becomes a common factor 
\begin{equation}
\int_{-\infty }^{\infty }dX\,\exp \left( -\frac{2\sin ^{2}\alpha }{l_{H}^{2}}%
\,X^{2}\right) =\sqrt{\frac{\pi }{2}}\frac{l_{H}}{\sin \alpha }.
\end{equation}
Pulling out this factor, we obtain 
\begin{eqnarray}
&<&|d|^{4}>_{0}=\sqrt{\frac{\pi }{2}}\frac{l_{H}}{b\sin \alpha }\left\{
\sum_{\text{even }M,N}\exp \left\{ 2\pi i\frac{b}{a}\cos \alpha \left[
-\left( \frac{M}{2}\right) ^{2}+\left( \frac{N}{2}\right) ^{2}\right] -\frac{%
b^{2}\sin ^{2}\alpha }{l_{H}^{2}}\left[ \left( \frac{M}{2}\right)
^{2}+\left( \frac{N}{2}\right) ^{2}\right] \right\} \right.  \nonumber \\
&&\left. +\sum_{\text{odd }M,N}\exp \left\{ 2\pi i\frac{b}{a}\cos \alpha
\left[ -\left( \frac{M}{2}\right) ^{2}+\left( \frac{N}{2}\right) ^{2}\right]
-\frac{b^{2}\sin ^{2}\alpha }{l_{H}^{2}}\left[ \left( \frac{M}{2}\right)
^{2}+\left( \frac{N}{2}\right) ^{2}\right] \right\} \right\}  \label{d40}
\end{eqnarray}
It is convenient to introduce the complex variable \cite{SaintJames} 
\begin{equation}
\zeta \equiv \frac{b}{a}\exp (i\alpha )\equiv \rho +i\sigma \text{.}
\label{zeta}
\end{equation}
The coefficients of $M^{2}$ and $N^{2}$ in Eq.(\ref{d40}) then become $-2\pi
i\zeta ^{*}/4$ and $2\pi i\zeta /4$. Finally, we obtain 
\begin{equation}
\beta _{A}^{0}=\sqrt{\sigma }\left\{ \sum_{n=-\infty }^{\infty }\left| \exp
(2\pi i\zeta n^{2})\right| ^{2}+\left| \sum_{n=-\infty }^{\infty }\exp
\left[ 2\pi i\zeta (n+\frac{1}{2})^{2}\right] \right| ^{2}\right\} .
\label{beta0}
\end{equation}
The above calculation can be straightforwardly extended to include the
perturbation of $\eta $. The relevant integral now becomes 
\[
\int_{-b\sin \alpha }^{0}dx\,\psi _{1}(x-nb\sin \alpha )\psi _{0}\,(x-m\sin
\alpha )\psi _{0}(x-n^{\prime }\sin \alpha )\psi _{0}(x-m^{\prime }\sin
\alpha )\,, 
\]
where $\psi _{1}$ is now given by Eq.(\ref{perturbedd}). The correction of $%
\beta _{A}$ in the first order of $\eta ^{\prime }$, after some tedious
algebra, is: 
\begin{eqnarray*}
\beta _{A}^{1} &=&\frac{\eta ^{\prime }}{4}\sqrt{\sigma }%
\mathop{\rm Re}%
\left\{ \exp (4i\varphi )\left[ \sum_{n^{\prime }}\exp (-2\pi i\zeta
^{*}n^{\prime 2})\right] \left[ \sum_{n}\exp (2\pi i\zeta n^{2})(64\pi
^{2}\sigma ^{2}n^{4}-48\pi \sigma n^{2}+3)\right] \right. \\
&&\left. +\left( n\rightarrow n+\frac{1}{2},n^{\prime }\rightarrow n^{\prime
}+\frac{1}{2}\right) \right\}
\end{eqnarray*}

\subsection{Minimization of the free energy and optimal vortex lattice
structure}

Having calculated the Abrikosov parameter $\beta _{A}$, one finds the vortex
structure by minimizing it with respect to $\varphi $, $\rho $, and $\sigma $%
. The minimization with respect to the angle $\varphi $ between the vortex
lattice and the crystal axes is easily done analytically. The general form
of $\beta _{A}$ is 
\begin{equation}
\beta _{A}(\varphi ,\rho ,\sigma )=\beta _{A}^{0}(\rho ,\sigma )+\eta \left[
e^{4i\varphi }\delta (\rho ,\sigma )+e^{-4i\varphi }\delta ^{*}(\rho ,\sigma
)\right] .  \label{fullbeta}
\end{equation}
Obviously the minimum of $\beta _{A}$ is achieved when 
\begin{equation}
\varphi =-{\rm \arg }[\delta (\rho ,\sigma )]/4\pm \pi /4.  \label{phi}
\end{equation}
The minimum of $\beta _{A}$ is 
\begin{equation}
\beta _{A}^{\min }(\rho ,\sigma )=\beta _{A}^{0}(\rho ,\sigma )-|\eta \delta
(\rho ,\sigma )|.  \label{betamin}
\end{equation}

The further minimization of $\beta _{A}^{\min }(\rho ,\sigma )$ is done
numerically. In Fig. 5, we show a plot of $\beta _{A}^{\min }(\rho ,\sigma )$
for $\eta ^{\prime }=0.0193$. Due to the fact that the same vortex lattice
might be represented by several sets of $(\rho ,\sigma )$, it is enough to
consider the region $0<\rho <1/2$, see discussion in \cite{SaintJames}. For
every $\eta ^{\prime }$, we observed that two degenerate minima appear. One
is at $\rho =1/2,\sigma =0.663$, and is clearly an rectangular body centered
lattice with $%
\alpha =53^{o}$. The corresponding $\varphi $ is zero. Therefore the vortex
lattice coincides with the crystal axes. The same result was claimed in \cite
{Berlinsky2}. The other minimum is at $\rho =0.275$ and $\sigma =0.961$ and $%
\alpha =74^{o}$, but with $\varphi $ equal to $37^{o}$. This corresponds to
the previous lattice rotated by $\pi /2$. To conclude, we observed rectangular
body centered
lattices only. The lowest energy state is doubly degenerate. It is
interesting to note that the fourfold symmetry of the underlying crystal is
not completely broken spontaneously by the static vortex lattice - rotations
of $\pi $ and reflections are symmetries of both rectangular lattices. The $\eta
^{\prime }$ dependence of $\alpha $ and $\beta _{A}^{\min }$ are plotted in
Fig. 6 and Fig. 7. We observed that there is a phase transition occured at $%
\eta ^{\prime }=0.0235$ where the lattice goes continuously from rectangular to
square.

Note that the calculations described in this section, unlike those for the
single vortex, are valid for arbitrary (not only large $\kappa $) type II
superconductor$.$ Therefore the results can be applied to non high $T_{c}$
materials as well. Using standard methods, one can take into account
variations of the magnetic field. We do not repeat here the standard
relations between the free energy and the Abrikosov parameter $\beta _{A}$ 
\cite{SaintJames}. One also can calculate corrections to the magnetization
curves in a standard way using the $\beta _{A}$ calculated here.

Despite the fact that general oblique lattices were considered for the first
time for d - wave, our numerical analysis shows that they have higher energy
than the rectangular body centered ones. Intuitively in the symmetric 
case this is
understandable because the rectangular lattices are more symmetric.
Although for the
s - wave superconductors this fact has been established \cite{Lasher}, for
rotationally non-symmetric superconductors this ''argument'' is not invalid.
We are not aware of any mathematical investigations of this question.
Moreover, when the vortex lattice starts moving, the rotational symmetry is
further explicitly broken. As we will see in next section, the general oblique
lattices nevertheless are not formed.

\section{The moving lattice solution of the time dependent Ginzburg-Landau
equation}

In this chapter we generalize the above procedure to find the structure for
a moving vortex lattice near $H_{c2}$ (the upper critical field itself being
a function of the current, as will be discussed in Subsection V.B). One can
consider the motion as caused by electric field ${\bf E}$. The vortex
lattice velocity is perpendicular to both electric and magnetic field
(assumed not to be tilted for simplicity and taken to be in the $+z$
direction): ${\bf E}=-{\bf v}\times {\bf B}.$ For a general direction of the
electric field the fourfold symmetry of the system is completely
(explicitly) broken. Only for several special directions, along the crystal
axes [1,0,0], [0,1,0] or along the diagonal lines [1,1,0] or [1,\={1},0] the
crystal symmetry is not broken completely, only reduced. Even for the simple 
$s$ - wave time dependent GL equations (TDGL) the problem of finding the
moving lattice solution is nontrivial. However there exists the ''Galilean
boost'' trick \cite{Hu} to solve the linearized (and sometimes even a
nonlinear problem for linear response\cite{Dorsey}) problem. As we will see
shortly, for the $d$ - wave equations, even the linearized equation does not
seem to possess a boosted static solution.

Technically the steps follow those of the static case. First we find a
complete set of solutions of the linearized equations using perturbation
theory in $\eta $. Then we impose the periodicity conditions to construct
the vortex lattice wave functions. It is more convenient to perform the
first step in the gauge aligned in the direction of the electric field,
while for the second step it is preferable to use a gauge aligned in the
direction of the vortex lattice. We will combine the two steps using gauge
transformation. After the wave function is found, it is straightforward to
apply the procedure described in the previous section to minimize the
Abrikosov's $\beta _{A}$.

\subsection{The perturbative solution of the linearized TDGL equations}

To simplify the presentation, we first assume that the direction of the
electric field is special: along the cryslalline $x$ (or [1,0,0]) direction.
In this case the vortices are moving in the negative $y$ direction of the
coordinate system. We will return to the general case afterwards. Now we
will construct the perturbative solution to the linearized TDGL Eq.(\ref
{TDGL}) which we now write in the following ''diffusion equation'' form:

\begin{equation}
\gamma \left( \frac{\partial }{\partial t}+ie^{*}\Phi \right) d=-\left( 
\frac{1}{2m_{d}}\Pi ^{2}-\alpha _{d}\right) d+\eta \left( \Pi _{y}^{2}-\Pi
_{x}^{2}\right) ^{2}d  \label{linearized TDGL}
\end{equation}

The vector potential we adopt here is the same as that in section IV, while
the electric potential can be chosen to be time and $y$ independent: 
\begin{equation}
\Phi =-vHx.  \label{electricpotential}
\end{equation}
In this gauge, the variables $t$ and $y$ trivially separate from $x$,

\[
d(x,y,t)=\exp (iky)\exp (-\omega t/\gamma )\psi (x), 
\]
where $\omega $ can have an imaginary part: $\omega =\omega _{R}+i\omega
_{I} $. The equation then reduces to a one dimensional ''Schroedinger type''
one (note there is an anti - Hermitian dissipation term) : 
\[
\left\{ \frac{1}{2m_{d}}\left[ \hat{p}_{x}^{2}+\left( k-\frac{x}{l_{H}^{2}}%
\right) ^{2}\right] -i\gamma ve^{*}Hx-\alpha _{d}-\eta \left( \Pi
_{y}^{2}-\Pi _{x}^{2}\right) ^{2}\right\} \psi (x)=\omega \psi (x) 
\]
Completing the square, rearranging the equation and noting that $\omega
_{I}=-ik\gamma v$ one obtains: 
\begin{eqnarray}
&&\left\{ \frac{1}{2m_{d}}\left[ -\frac{d^{2}}{dx^{2}}+\frac{1}{l_{H}^{4}}%
(x-x_{0}-igl_{H})^{2}\right] -\eta \,(\Pi _{y}^{2}-\Pi _{x}^{2})^{2}-\alpha
_{d}+\frac{1}{2}\gamma ^{2}m_{d}v^{2}\right\} \psi (x) \\
&\equiv &\left[ \left( \hat{K}-\eta \hat{V}\right) -\alpha _{d}+\frac{1}{2}%
\gamma ^{2}mv^{2}\right] \psi (x)=\omega _{R}\psi (x).  \label{eigeneq}
\end{eqnarray}
where a dimensionless quantity $g$ is defined by $g\equiv \gamma
m_{d}\,v\,l_{H}$ and $x_{0}=kl_{H}^{2}$. \ The parameter $\alpha _{d}$
should be adjusted (i.e., changing the temperature) such that the lowest
eigenvalue $\omega _{R}$ becomes zero, otherwise one gets runaway solutions.
This is nothing but the $H_{c2}$ condition generalized to include arbitrary
electric field. The operator $\hat{K}$ defined in Eq.\ref{eigeneq}) is
simply $\tilde{K}\equiv \Pi ^{2}/2m_{d}$ with $\tilde{x}\equiv
(x-x_{0})/l_{H}$ shifted by an imaginary amount $-ig$, so we can write it as
: 
\begin{equation}
\hat{K}=\exp (g\,l_{H}\hat{p}_{x})\,\tilde{K}\exp (-g\,l_{H}\,\hat{p}_{x}).
\label{shiftedh}
\end{equation}

The perturbation theory to Eq.(\ref{eigeneq}) is most conveniently performed
on the shifted $\psi $ field defined by: 
\begin{equation}
\psi (\tilde{x})\equiv \exp (g\,l_{H}\hat{p}_{x})\,\tilde{\psi}(\tilde{x})=%
\tilde{\psi}(\tilde{x}-ig).  \label{shiftedfield}
\end{equation}
The transformed Hamiltonian is: 
\[
\tilde{H}\equiv \exp (-g\,l_{H}\hat{p}_{x})\left( \hat{K}-\eta \hat{V}%
\right) \exp (gl_{H}\,\hat{p}_{x}) 
\]
Going to the creation and annihilation operators $\hat{a}^{\dagger }$ and $%
\hat{a}$ representation, Eq.(\ref{eigeneq}) becomes 
\begin{equation}
\left[ \hat{a}^{\dagger }\hat{a}+\frac{1}{2}-\eta ^{\prime }\,\tilde{V}(\hat{%
a},\hat{a}^{\dagger })\right] \tilde{\psi}(x)=\frac{m_{d}}{e^{*}H}\left(
\omega _{R}+\alpha _{d}-\frac{1}{2}\gamma ^{2}m_{d}v^{2}\right) \,\tilde{\psi%
}(x)\equiv \xi \tilde{\psi}(x).  \label{shiftH}
\end{equation}
Here 
\begin{equation}
\tilde{V}(\hat{a},\hat{a}^{\dagger })=\exp \left[ -i\frac{g}{\sqrt{2}}(\hat{a%
}^{\dagger }-\hat{a})\right] \left( \hat{a}^{\dagger 2}+\hat{a}^{2}\right)
^{2}\exp \left[ i\frac{g}{\sqrt{2}}(\hat{a}^{\dagger }-\hat{a})\right] .
\label{shiftedV}
\end{equation}
The ''potential energy''can be further simplified using the identities: 
\begin{eqnarray}
\exp \left[ -i\frac{g}{\sqrt{2}}(\hat{a}^{\dagger }-\hat{a})\right] \,\hat{a}%
\exp \left[ i\frac{g}{\sqrt{2}}(\hat{a}^{\dagger }-\hat{a})\right] &=&\hat{a}%
+i\frac{g}{\sqrt{2}}, \\
\exp \left[ -i\frac{g}{\sqrt{2}}(\hat{a}^{\dagger }-\hat{a})\right] \,\hat{a}%
^{\dagger }\exp \left[ i\frac{g}{\sqrt{2}}(\hat{a}^{\dagger }-\hat{a}%
)\right] &=&\hat{a}^{\dagger }+i\frac{g}{\sqrt{2}}.
\end{eqnarray}
It is helpful to note that the state resulting from the action of the
shifting operator on $|0\rangle $ is a coherent state 
\begin{equation}
|-i\frac{g}{\sqrt{2}}\rangle \equiv \exp \left[ -i\frac{g}{\sqrt{2}}(\hat{a}%
^{\dagger }-\hat{a})\right] |0\rangle .
\end{equation}
The correction to the eigenvalue $\xi $ (used later to find the phase
transition boundary) to the first order $\eta \,$ is then easily found:

\begin{equation}
\xi =\frac{1}{2}-\eta ^{\prime }\,(g^{4}-2g^{2}+2)  \label{ksistat}
\end{equation}
$.$To the first order in $\eta $, the perturbed ground state is given by: 
\begin{equation}
\tilde{\psi}=|0\rangle +\sum_{n=1}^{4}\frac{\eta ^{\prime }}{n}\,|n\rangle
\langle n|\left[ (\hat{a}^{\dagger }+i\frac{g}{\sqrt{2}})^{2}+(\hat{a}+i%
\frac{g}{\sqrt{2}})^{2}\right] ^{2}|0\rangle \equiv |0\rangle +\eta ^{\prime
}\sum_{n=1}^{4}c_{n}\,|n\rangle .
\end{equation}
where 
\begin{equation}
\left\{ 
\begin{array}{l}
c_{1}=-2\sqrt{2}ig(g^{2}-1) \\ 
c_{2}=-2\sqrt{2}g^{2} \\ 
c_{3}=\frac{4\sqrt{3}}{3}ig \\ 
c_{4}=\frac{\sqrt{6}}{2}
\end{array}
\right.
\end{equation}
The solution of the original linearized TDGL equation Eq.(\ref{linearized
TDGL}) is simply the shifted $\psi (x)$ together with other factors: 
\begin{eqnarray}
d(x,y,t) &=&\exp [\,ik(y+vt)]\,\times \exp \left[ -\frac{1}{2l_{H}^{2}}\left( x-kl_{H}^{2}-igl_{H}\right) ^{2}\right] \times  \nonumber \\
&&
\left( \frac{1}{\pi  l_{H}^{2}}\right)^{1/4}
\left[ 1+\eta \sum_{n=1}^{4}c_{n}\frac{1}{\sqrt{2^{n}n!}}
H_{n}\left( \frac{x}{l_{H}}-kl_{H}-ig\right) \right]
\label{non-rotated TDGL solution}
\end{eqnarray}
%

The solution we constructed is restricted to the case when \vspace{1pt} the
direction of the electric field is along the crystalline $x$ direction. \
Now we generalize the calculation to arbitrary direction of the electric
field. In the coordinate system fixed by the vortex lattice (which we will
use for construction of the vortex lattice solution) the general vortex
velocity is: $v_{x}=v\sin \theta ,v_{y}=-v\cos \theta ,$ while the electric
field is $E_{x}=E\cos \theta ,E_{y}=E\sin \theta .$ The angle between the
crystal [1,0,0] axis and the electric field will be therefore $\Theta
=\theta -\varphi $ (see Fig. 4). The calculation is just a little bit more
complicated. It still will be convenient to choose a coordinate system in
which the direction of the electric field and that of the $x$ axis coincide.
The perturbed Hamiltonian then becomes:

\[
\tilde{V}(\hat{a},\hat{a}^{\dagger })=\exp \left[ -i\frac{g}{\sqrt{2}}(\hat{a%
}^{\dagger }-\hat{a})\right] \left( e^{-2i(\theta -\varphi )}\hat{a}%
^{\dagger 2}+e^{2i(\theta -\varphi )}\hat{a}^{2}\right) ^{2}\exp \left[ i 
\frac{g}{\sqrt{2}}(\hat{a}^{\dagger }-\hat{a})\right] 
\]

The corrected solution has the same form as Eq.(\ref{non-rotated TDGL
solution}), with the coefficients $c_{n}$ which now depend on the angle $%
(\theta -\varphi )$, as follows:\newline
\begin{equation}
\left\{ 
\begin{array}{l}
c_{1}=-\sqrt{2}ig\left[ \left( 1+e^{-4i\left( \theta -\varphi \right)
}\right) g^{2}-2\right] \\ 
c_{2}=-\frac{\sqrt{2}}{2}\left( 1+3e^{-4i\left( \theta -\varphi \right)
}\right) g^{2} \\ 
c_{3}=\frac{4\sqrt{3}}{3}ige^{-4i\left( \theta -\varphi \right) } \\ 
c_{4}=\frac{\sqrt{6}}{2}e^{-4i\left( \theta -\varphi \right) }
\end{array}
\right.  \label{smallc}
\end{equation}

The corresponding eigenvalue becomes 
\begin{equation}  \label{ksiangle}
\xi =\frac{1}{2}-\eta ^{\prime }\,\left[ \frac{1+\cos 4(\theta -\varphi )}{2}%
g^{4}-2g^{2}+2\right] .
\end{equation}

\subsection{\protect\vspace{1pt}\protect\vspace{1pt}Dependence of $H_{c2}$
on the electric field}

\vspace{1pt}In the simpler case of electric field parallel to one of the
crystal axes, from Eq.(\ref{ksistat}) the new phase boundary equation
follows: 
\[
H_{c2}=H_{c2}^{0}+\eta H_{c2}^{1}=\frac{m_{d}}{e^{*}}(2\alpha _{d}-\gamma
^{2}m_{d}v^{2})+\frac{2\eta m_{d}^{3}}{e^{*}}\left( 5m_{d}^{2}\gamma
^{4}v^{4}-12m_{d}\gamma ^{2}v^{2}\alpha _{d}+8\alpha _{d}^{2}\right) 
\]
where the second term is a perturbation. All the temperature dependence (at
least within the confines of the simple Ginzburg - Landau assumption) is
contained inside $\alpha _{d}=\alpha ^{\prime }(T_{c}-T)$. The phase
transition line is therefore still quadratic in $T,$

\begin{equation}
H_{c2}(T)=h_{0}+h_{1}(T_{c}-T)+h_{2}(T_{c}-T)^{2}  \label{line}
\end{equation}
but first two coefficients have a nontrivial dependance on velocity $v$: 
\begin{eqnarray}
h_{0} &=&\frac{m_{d}^{2}}{e^{*}}\left( -1+10\eta m_{d}^{3}\gamma
^{2}v^{2}\right) \gamma ^{2}v^{2}  \nonumber \\
h_{1} &=&2\alpha ^{\prime }\frac{m_{d}}{e^{*}}\left( 1-12\eta
m_{d}^{3}\gamma ^{2}v^{2}\right)  \label{line1} \\
h_{2} &=&16\alpha ^{\prime 2}\eta \frac{m_{d}^{3}}{e^{*}}  \nonumber
\end{eqnarray}
Note that the curvature hasn't changed compared to the static case, but we
have two new effects. First of all, the electric field (or, equivalently,
electric current) shifts $H_{c2}$ by a negative constant (proportional to $%
E^{2\text{ }}$) to a lower value. This is expected. Secondly, although the
curvature $h_{2}$ doesn't change compared with the static case, the slope $%
h_{1}$ acquires a negative contribution proportional to $E^{2\text{ }}$.

In the general case of arbitrary orientation of the electric field, with
respect to the crystal lattice, only the coefficient $h_{0}$ needs to be
changed to:

\begin{equation}
h_{0}=\frac{m_{d}^{2}}{e^{*}}\left\{ -1+\left[ 9+\cos 4(\theta -\varphi
)\right] \eta m_{d}^{3}\gamma ^{2}v^{2}\right\} \gamma ^{2}v^{2}
\label{line2}
\end{equation}
We got the interesting result that the shift in $H_{c2}$ due to the electric
field actually depends on the direction of the electic field relative to the
crystal lattice. This result should be checked experimentally.

In the s - wave case (or $\eta =0)$ the boundary was first found and
discussed in ref. .\cite{Hu}. There are a couple of peculiarities associated
with it like the existence of a metastable normal state and the unstable
superconductive state. The same applies to the present case. As far as we
know, these peculiarities haven't been directly observed in low $T_{c}$
materials. It would be interesting to reconsider this question for the high $%
T_{c}$ materials. Note also that the phase transition is not a usual one
(second order which probably turns to weakly first order due to
fluctuations). In the presence of flux flow the two phases are stationary
states rather than states in thermal equilibrium. There exist therefore a
phase diagram in the space containing the current as an external parameter
(both magnitude and direction).

\subsection{\protect\smallskip Construction of the moving vortex lattice}

Now we would like to follow a procedure similar to that described in Section
IV.C for the static case to construct a vortex lattice solution. It turns
out not to be a straightforward generalization. In earlier sections, we used
the gauge freedom to make both the scalar and the vector potentials
independent of $y$ and $t$. This allows for separation of variables. The
fact that $y$ variable factored into the form $\exp (iky)$ helped us
implement the periodicity in the $y$ direction (with discrete values of $k$%
). However, in general, the vortex lattice will not be periodic along this
special direction. To construct this general periodic solution, one has to
solve a very complicated periodicity constraint equation for the
coefficients $C_{k},$ where $k$ is now a continuous index.

In the static vortex lattice case, we used the gauge freedom to align the
vector potential to the vortex lattice. This choice allows us to solve the
constraint equation on $C_{k}$ easily since we already had periodicity along
the $y-$axis built in. This reduced the problem to a discrete one.
Furthermore, only a few $k_{n}$'s were coupled, and it turned out to be
solvable, at least for $p=1.$ This is not the case for the moving vortex
lattice. In Subsection A, for the problem with electric field and time
dependence, we used the gauge freedom to align the vector potential with
respect to the electric field in order to find the general solution of the
perturbed Hamiltonian. Now, when we have to use this general solution to
construct the periodic solution we encounter the problem that we cannot use
the gauge freedom to simultaneously simplify {\it both} problems.
Fortunately, in the unperturbed (s-wave) case a simple Anzatz for the
construction of moving vortex lattice solution exists. This works for the
linearized TDGL equation with arbitrary direction of the electric field \cite
{Hu}. We shall use this observation to guide us in obtaining the periodic
solution for the moving vortex lattice in the presence of perturbation. The
solution can be explicitly checked to satisfy TDGL equation and the
periodicity constraints.

In subsection A, we were forced to choose an axial gauge as in Eq.(\ref
{electricpotential}) (which will be referred to later as the gauge I),

\begin{eqnarray}
{\bf A}^{I}(x^{\prime },y^{\prime },t) &=&Hx^{\prime }\hat{y}^{\prime } 
\nonumber \\
\Phi ^{I}(x^{\prime },y^{\prime },t) &=&{\bf v}\cdot {\bf A}^{I}{\bf =-}%
vHx^{\prime },
\end{eqnarray}
For later convenience, we use $x^{\prime },y^{\prime }$ to represent the
coordinate in which the electric field is along $x^{\prime }$ direction,
while $x,y$ is the coordinate in which the vortex lattice is alligned to the 
$y$ direction (see Fig.4). The relation is:

\[
\left\{ 
\begin{tabular}{l}
$x^{\prime }=x\cos \theta +y\sin \theta $ \\ 
$y^{\prime }=-x\sin \theta +y\cos \theta $%
\end{tabular}
\right. 
\]

Fortunately we can transform our solution to a gauge in which the
periodicity is manifest and the standard procedure works (referred as gauge
II): 
\begin{eqnarray}
{\bf A}^{II}(x,y,t) &=&H\left( x-v_{x}t\right) \hat{y}+\gamma \frac{m_{d}}{%
e^{*}}{\bf v}\times \hat{z}  \nonumber \\
\Phi ^{II}(x,y,t) &=&{\bf v}\cdot {\bf A}^{B}=v_{y}H\left( x-v_{x}t\right)
\end{eqnarray}
where $v_{x}=v\sin \theta ,v_{y}=-v\cos \theta .$ The gauge transformation
between two is determined by a phase $\chi (x,y,t)$ satisfying 
\begin{eqnarray*}
\nabla \chi &=&{\bf A}^{I}-{\bf A}^{II} \\
-\partial _{t}\chi &=&\Phi ^{I}-\Phi ^{II}
\end{eqnarray*}
One of the solutions is:

\begin{equation}
\chi =\frac{\gamma m_dv}{e^{*}}x^{\prime }+\frac H2\sin \theta \cos \theta
\left[ \left( y^{\prime }+vt\right) ^2-x^{^{\prime }2}\right] +H\sin
{}^2\theta \left[ x^{\prime }\left( y^{\prime }+vt\right) \right]
\end{equation}
.

In this gauge unperturbed lattice can be easily formed using ''boosted''
solutions,

\begin{equation}
\Psi _{n}^{II}\left( x,y,t\right) =\frac{1}{\sqrt{L}}\left( \frac{1}{\pi
l_{H}^{2}}\right) ^{\frac{1}{4}}\exp \left[ ik_{n}\left( y-v_{y}t\right)
\right] \exp \left[ -\frac{1}{2l_{H}^{2}}\left(
x-v_{x}t-k_{n}l_{H}^{2}\right) ^{2}\right]
\end{equation}
with standard coefficients: $\Psi ^{II}=\sum_{n}C_{n}\Psi _{n}^{II}$. These
elementary solutions are linearly related to the unperturbed normalized
eigenfunctions in the gauge I found in Subsection V.A, 
\begin{equation}
\Psi _{k}^{I}\left( x^{\prime },y^{\prime },t\right) =\frac{1}{\sqrt{L}}%
\left( \frac{1}{\pi l_{H}^{2}}\right) ^{\frac{1}{4}}\exp \left( -\frac{g^{2}%
}{2}\right) \exp \left[ ik\left( y^{\prime }+vt\right) \right] \exp \left[ -%
\frac{1}{2l_{H}^{2}}\left( x^{\prime }-igl_{H}-kl_{H}^{2}\right) ^{2}\right]
\end{equation}
after the gauge transformation is performed: 
\[
e^{ie^{*}\chi }\Psi _{n}^{II}=\frac{L}{2\pi }\int_{-\infty }^{\infty
}dkB_{nk}\Psi _{k}^{I} 
\]
Note that gauge transformation and hence quantities $B_{nk}$ are in general
time dependent. However, we do not need to keep track of the time dependence
here. The reason is that the GL equation we are solving with or without the
perturbation is time translation invariant. While keeping track of the time
dependence of gauge non-invariant quantities $B_{nk}$ and the wave functions
is complicated, because in this case we would have to use a time dependent
gauge choice, the gauge invariant quantities such as $\beta _{A}$ are
automatically time independent. Therefore, to simplify the calculation, we
can set $t=0$.

To find the coefficients $B_{nk}$ two gaussian integrations should be
performed: 
\begin{eqnarray}
B_{nk} &=&\int dx^{\prime }dy^{\prime }\left[ \Psi _{k}^{I*}\left( x^{\prime
},y^{\prime },0\right) e^{ie^{*}\chi \left( x^{\prime },y^{\prime },0\right)
}\Psi _{n}^{II}\left( x,y,0\right) \right] = \\
&&\frac{\sqrt{\pi }l_{H}}{L}\frac{1}{\sqrt{ie^{i\theta }\sin \theta }}\exp
\left[ -\frac{1}{2}\frac{\cos \theta }{\sin \theta }\left(
k^{2}+k_{n}^{2}\right) l_{H}^{2}+ikl_{H}^{2}\left( \frac{k_{n}}{\sin \theta }%
+\gamma m_{d}v\right) \right]
\end{eqnarray}

We have already found the first order correction to the wave function is
gauge I. The corresponding expression in the gauge II is: 
\[
e^{ie^{*}\chi }\delta \Psi _{n}^{II}=\frac{L}{2\pi }\int_{-\infty }^{\infty
}dk\;B_{nk}\delta \Psi _{k}^{I} 
\]
where 
\begin{eqnarray}
\delta \Psi _{k}^{I} &=&\exp [\,ik(y^{\prime }+vt)]\,\times \exp \left[ -%
\frac{1}{2l_{H}^{2}}\left( x^{\prime }-kl_{H}^{2}-igl_{H}^{2}\right)
^{2}\right] \times  \nonumber \\
&&\eta ^{\prime }\sum_{n=1}^{4}c_{n}\frac{1}{\sqrt{2^{n}n!}}\left( \frac{H}{%
\pi }\right) ^{1/4}H_{n}\left( \frac{x^{\prime }}{l_{H}}-kl_{H}-ig\right)
\end{eqnarray}
as was shown in the previous section.

It turns out that after a lengthy calculation, the correction to the wave
function in gauge II is amazingly simple:

\begin{eqnarray}
\delta \Psi _{n}^{II}\left( x,y\right) &=&e^{-ie^{*}\chi }\frac{L}{2\pi }%
\int_{-\infty }^{\infty }dk\;B_{nk}\delta \Psi _{k}^{I}  \nonumber \\
&=&\frac{1}{\sqrt{L}}\left( \frac{1}{\pi l_{H}^{2}}\right) ^{1/4}\exp \left(
\,ik_{n}y\right) \exp \left[ -\frac{1}{2l_{H}^{2}}\left(
x-k_{n}l_{H}^{2}\right) ^{2}\right]  \nonumber \\
&&\times \eta ^{\prime }\sum_{m=1}^{4}c_{m}\frac{e^{im\theta }}{\sqrt{2^{m}m!%
}}H_{m}\left( \frac{x}{l_{H}}-k_{n}l_{H}-ige^{-i\theta }\right) .
\end{eqnarray}
An important observation is that the corrected moving lattice solution at $%
t=0$

\begin{eqnarray}
\Psi ^{II}(x,y) &=&\sum_{n=-\infty }^{\infty }C_{n}\left[ \Psi
_{n}^{II}+\delta \Psi _{n}^{II}\right]  \nonumber \\
&=&\sum_{n=-\infty }^{\infty }C_{n}\frac{1}{\sqrt{L}}\left( \frac{H}{\pi }%
\right) ^{1/4}\exp \left( \,ik_{n}y\right) \exp \left[ -\frac{1}{2l_{H}^{2}}%
\left( x-k_{n}l_{H}^{2}\right) ^{2}\right] \times  \nonumber \\
&&\left[ 1+\eta ^{\prime }\sum_{m=1}^{4}c_{m}\frac{e^{im\theta }}{\sqrt{%
2^{m}m!}}H_{m}\left( \frac{x}{l_{H}}-k_{n}l_{H}-ige^{-i\theta }\right)
\right]
\end{eqnarray}
where $k_{n}=\frac{2\pi n}{a}$, and $C_{n}$ again given by Eq.(\ref{C_n}) is
still invariant under the vortex lattice symmetries. From now on we work
exclusively in gauge II. Physically this is understood as follows. Our
perturbation commutes with all the translations generators, in particular
with the translations defining the lattice structure. Therefore if the
unperturbed function has certain lattice translation symmetry, the perturbed
one will have as well.

\subsection{The structure and magnetization of the moving lattice}

The standard Abrikosov's procedure to develope an approximation for small
order parameter around $H_{c2}$ can be applied also in the flux flow case
(see \cite{Hu}). This time however the minimization of the Abrikosov
parameter $\beta _{A}$ does not correspond to minimization of energy, but
rather to smallest deviation from the exact solution of TDGL equation. The
''derivation'' closely follows the static one. Using the expression for the
vortex lattice solution found in the previous subsection, the correction
term in expansion of the Abrikosov parameter $\beta _{A}$ in $\eta ^{\prime
},$

\[
\beta _{A}=\beta _{A}^{0}+\eta ^{\prime }\beta _{A}^{1} 
\]
is: 
\begin{eqnarray}
\beta _{A}^{1} &=&\frac{\sqrt{\sigma }}{4}%
\mathop{\rm Re}%
\left\{ \left[ \sum_{n^{\prime }}\exp (-2\pi i\zeta ^{*}n^{\prime 2})\right]
\left[ \sum_{n}\exp (2\pi i\zeta n^{2})G(n)\right] \right. +  \nonumber \\
&&\left. \left( n\rightarrow n+\frac{1}{2},n^{\prime }\rightarrow n^{\prime
}+\frac{1}{2}\right) \right\} .  \label{beta1}
\end{eqnarray}
Here the function $G(n)$ is defined by 
\begin{eqnarray}
G(n) &=&e^{4i\varphi }(64\pi ^{2}\sigma ^{2}n^{4}-48\pi \sigma n^{2}+3) 
\nonumber \\
&&-8e^{2i\varphi }g^{2}\cos 2\Theta (8\pi \sigma n^{2}-1)  \label{G}
\end{eqnarray}
where $\Theta \equiv \theta -\varphi $ is the angle between the electric
field and the crystal lattice. One immediately observes a surprising fact -
the dependance on the angle $\Theta $ and velocity $v$ is only via the
combination $g^{2}\cos 2\Theta $ where $g\equiv \gamma m_{d}\,v\,l_{H}$. It
factors out as:

\begin{equation}
\beta _{A}(\varphi ,\rho ,\sigma )\equiv \beta _{A}^{0}(\rho ,\sigma )+\eta
^{\prime }%
\mathop{\rm Re}%
\left[ e^{4i\varphi }\delta (\rho ,\sigma )+g^{2}\cos 2\Theta e^{2i\varphi
}\delta ^{\prime }(\rho ,\sigma )\right] .  \label{oblique}
\end{equation}
For example, the resulting lattice for $\Theta =\pi /4$ and arbitrary $g$
will be the same as without electric field at all! Also apparent complete
breaking of the rotational symmetry by general direction of the electric
field is not felt by $\beta _{A}.$ Indeed, lattice for some arbitrary $%
\Theta $ and $g^{2}$ is the same as for $\Theta =0$ and $g^{2\prime
}=g^{2}\cos 2\Theta $. The fourfold symmetry has been reduced however. These
results are nontrivial and can be checked experimentally. This degeneracy in
velocity and $\Theta $ in the determination of vortex lattice may be a
reflection of some dynamical symmetries which we have so far failed to see
yet.

We get the $e^{\pm 2i\varphi }$ harmonics in Eq.(\ref{oblique}) in addition
to the fourth harmonic that appeared in the static case. The minimization
with respect to $\varphi $ still can be done analytically, although the
algebraic equation in this case is quartic. For fixed $\eta ^{\prime }$ and $%
g^{2}\cos 2\Theta ,$ the minimization with respect to $\rho $ and $\sigma $
was performed numerically and we again obtain only rectangular body centered
lattices aligned
either to the crystalline axes. The angle $\alpha $ turns out to be only
weakly dependent on the combination $g^{2}\cos 2\Theta .$ For example, for
positive $\eta ^{\prime }=0.015$ , we obtain $\alpha =\alpha
(g=0)+1.0g^{2}|\cos 2\Theta |$ (in degrees) where $\alpha (g=0)=69.3^{\circ }
$. The Abrikosov $\beta _{A}$ is simply related to the slope of the
magnetization curve 
\begin{equation}
4\pi \frac{dM}{dH}=\frac{1}{(2\kappa ^{2}-1)\beta _{A}}  \label{magn}
\end{equation}
as well as to other thermodynamic quantities. All of them therefore exhibit
very simple dependence on the velocity ${\bf v}$ , or, equivalently, on the
current ${\bf J}$.

The fact that the optimal lattice is rectangular body centered is a bit
mysterious. Rotational
symmetry is completely broken by both the electric field and by the
underlying crystal lattice. It is not easy to attribute the advantage of
this lattice structure to some simple physical origin. It might be that it
is a consequence of using the Abrikosov approximation, and therefore beyond
this approximation the lattices might not be rectangular. Note also that it was
also surprising that $\beta _{A}$ was independent of the orientation of the
electric field even in the s - wave calculation \cite{Hu}. As far as we
know, the preferred orientation of the moving lattice has not been observed
in either low $T_{c}$ or high $T_{c}$ type II superconductors.

\section{The non-linear conductivity near $H_{c2}$}

In this section we consider the dissipation in vortex cores due to flux
flow. As it is well known, the fourfold symmetry forces the conductivity
tensor $\sigma _{ij},$defined by $J_{i}=\sigma _{ij}E_{j}$, to be
rotationally symmetric. Namely, $\sigma _{ij}=\sigma \delta _{ij}+\sigma
^{H}\varepsilon _{ij}.$ Here $\sigma $ is the usual (Ohmic) conductivity, $%
\sigma ^{H}$ is the Hall conductivity and $\varepsilon _{ij}$ is the
antisymmetric tensor. The additional term in the free energy corrects the
values of $\sigma $ and $\sigma ^{H},$ but the correction is of the order $%
\eta $ and therefore small. So, to see anisotropy, we definitely would like
to go beyond linear response. This has been done for simple s - wave TDGL 
\cite{Hu} near $H_{c2\text{ }}$. We will neglect pinning and consider motion
of a very large bundle. While there is a normal component of the
conductivity, here we will concentrate on the contribution of the
supercurrent only. For the discussion of the relative contribution of the
two see \cite{Hu}.

\subsection{Condensate for the moving lattice}

To calculate the transport properties due to flux flow, we have to compute
two quantities. The first is the expression for the electric current and
will be obtained in the next subsection. The second is the normalization
factor for the d - wave order parameter. In the previous sections we only
needed Abrikosov's $\beta _{A}$ which is insensitive to the overall scale of
the condensate, but now we need to calculate it.

\[
d=N\sum C_{n}(\Psi _{n}+\eta ^{\prime }\delta \Psi _{n})\equiv N(\Psi +\eta
^{\prime }\delta \Psi )\cong (N_{0}+\eta ^{\prime }N_{1})(\Psi +\eta
^{\prime }\delta \Psi ) 
\]
Here $C_{n},\Psi _{n}$ and $\delta \Psi _{n}$ have been calculated in the
previous section, $N$ is the normalization and $N_{0}$ is the normalization
of $\Psi $ . The calculation is standard. We again expand it to first order
in $\eta ^{\prime }$. The normalization is determined from the minimization
of the free energy as

\[
<d^{*}d>=\frac{\alpha _{d}}{2\beta }\frac{1}{\beta _{A}} 
\]
where $<...>$ denote the space average, $\alpha _{d}$ and $\beta $ are
coefficients of the GL equation. The Abrikosov parameter $\beta _{A}$ has
its own $\eta ^{\prime }$ expansion: $\beta _{A}=\beta _{A}^{0}+\eta
^{\prime }\beta _{A}^{1}$ calculated in Section VI.C. Combining the two one
obtains

\[
N^{2}\cong N_{0}^{2}(1+\eta ^{\prime }\frac{2N_{1}}{N_{0}})=\frac{\alpha _{d}%
}{2\beta }\frac{1}{\beta _{A}^{0}<\Psi ^{*}\Psi >}\left[ 1-\eta ^{\prime
}\left( \frac{\beta _{A}^{1}}{\beta _{A}^{0}}+\frac{2%
\mathop{\rm Re}%
<\Psi ^{*}\delta \Psi >}{<\Psi ^{*}\Psi >}\right) \right] 
\]
which will be used to calculate the current.

\subsection{The direct and the Hall currents}

The anisotropy term in the free energy Eq.(\ref{fe}) contains four covariant
derivatives and consequently the electric current, in addition to the usual
expression, contains additional terms. The leading order current is given by

\[
{\bf J}^{a}=\frac{e^{*}}{2m_{d}}\langle d^{*}({\bf \Pi }d)+({\bf \Pi }%
d)^{*}d\rangle 
\]
The anisotropic perturbation to the current is

\begin{eqnarray}
{\bf J}^{b}(d) &=&e^{*}\eta {\bf \hat{x}}^{\prime \prime }\left\langle
\left[ \left( \Pi _{y}^{\prime \prime 2}-\Pi _{x}^{\prime \prime 2}\right)
d\right] ^{*}\left( \Pi _{x}^{\prime \prime }d\right) +\left[ \Pi
_{x}^{\prime \prime }\left( \Pi _{y}^{\prime \prime 2}-\Pi _{x}^{\prime
\prime 2}\right) d\right] ^{*}d+c.c.\right\rangle  \nonumber \\
&&-e^{*}\eta {\bf \hat{y}}^{\prime \prime }\left\langle \left[ \left( \Pi
_{y}^{\prime \prime 2}-\Pi _{x}^{\prime \prime 2}\right) d\right] ^{*}\left(
\Pi _{y}^{\prime \prime }d\right) +\left[ \Pi _{y}^{\prime \prime }\left(
\Pi _{y}^{\prime \prime 2}-\Pi _{x}^{\prime \prime 2}\right) d\right]
^{*}d+c.c.\right\rangle
\end{eqnarray}
where $\Pi _{x}^{\prime \prime }=\cos \varphi \Pi _{x}+\sin \varphi \Pi _{y%
\text{, }}\Pi _{y}^{\prime \prime }=\cos \varphi \Pi _{y}-\sin \varphi \Pi
_{x,}$ and ${\bf \hat{x}}^{\prime \prime }=\cos \varphi {\bf \hat{x}+\sin
\varphi \hat{y}}$, ${\bf \hat{y}}^{\prime \prime }=\cos \varphi {\bf \hat{y}%
-\sin \varphi \hat{x}}$ (See Fig. 4).

Substituting the condensate $d=N\left( \Psi +\eta ^{\prime }\delta \Psi
\right) $ with $\Psi $ and $\delta \Psi $ determined in the previous
subsection, one obtains the expansion of ${\bf J}^{a}$ to the first order in 
$\eta ^{\prime }$:

\[
\begin{array}{c}
{\bf J}^{a}{\bf \cong }N^{2}\left( \frac{e^{*}}{2m_{d}}\right) \left\{
\left[ \left\langle \Psi ^{*}{\bf \Pi }\Psi \right\rangle +\left\langle \Psi 
{\bf \Pi }\Psi ^{*}\right\rangle \right] +\right. \\ 
\left. \eta ^{\prime }\left[ \left\langle \Psi ^{*}{\bf \Pi }\delta \Psi
\right\rangle +\left\langle \delta \Psi ^{*}{\bf \Pi }\Psi \right\rangle
+\left\langle \delta \Psi {\bf \Pi }\Psi ^{*}\right\rangle +\left\langle
\Psi {\bf \Pi \delta }\Psi ^{*}\right\rangle \right] \right\} \\ 
\equiv {\bf j}+\eta ^{\prime }\left( \delta {\bf j}_{1}+\delta {\bf j}%
_{2}\right)
\end{array}
\]
where $\delta {\bf j}_{1}$ comes from the correction to $N^{2}$, and $\delta 
{\bf j}_{2}$ contains the correction to the wave function $\delta \Psi $. 
\begin{equation}
{\bf j=}N_{0}^{2}\left( \frac{e^{*}}{2m_{d}}\right) \left[ \left\langle \Psi
^{*}{\bf \Pi }\Psi \right\rangle +\left\langle \Psi {\bf \Pi }\Psi
^{*}\right\rangle \right] =\left( \frac{e^{*}\alpha _{d}}{4\beta m_{d}}\frac{%
1}{\beta _{A}^{0}}\right) \frac{2%
\mathop{\rm Re}%
\left\langle \Psi ^{*}{\bf \Pi }\Psi \right\rangle }{<\Psi ^{*}\Psi >}
\end{equation}
\[
\delta {\bf j}_{1}==-\left( \frac{\beta _{A}^{1}}{\beta _{A}^{0}}+\frac{2%
\mathop{\rm Re}%
<\Psi ^{*}\delta \Psi >}{<\Psi ^{*}\Psi >}\right) {\bf j} 
\]
\begin{equation}
\delta {\bf j}_{2}=\left( \frac{e^{*}\alpha _{d}}{4\beta m_{d}}\frac{1}{%
\beta _{A}^{0}}\right) \frac{4%
\mathop{\rm Re}%
\left\langle \delta \Psi ^{*}\Pi \Psi \right\rangle }{<\Psi ^{*}\Psi >}
\end{equation}
The expansion to first order in $\eta ^{\prime }$ of ${\bf J}^{b}$ is:

\[
{\bf J}^{b}(N\Psi )\simeq N_{0}^{2}{\bf J}^{b}(\Psi )\equiv \eta ^{\prime
}\delta {\bf j}_{3} 
\]
\begin{eqnarray*}
\delta {\bf j}_{3} &=&\left( \frac{e^{*}\alpha _{d}}{2\beta }\frac{1}{\beta
_{A}^{0}}\right) \left( \frac{l_{H}^{2}}{m_{d}}\right) \times \\
&&\left\{ {\bf \hat{x}}^{\prime \prime }\frac{\left\langle \left[ \left( \Pi
_{y}^{\prime \prime 2}-\Pi _{x}^{\prime \prime 2}\right) \Psi \right]
^{*}\left( \Pi _{x}^{\prime \prime }\Psi \right) +\left[ \Pi _{x}^{\prime
\prime }\left( \Pi _{y}^{\prime \prime 2}-\Pi _{x}^{\prime \prime 2}\right)
\Psi \right] ^{*}\Psi +c.c.\right\rangle }{<\Psi ^{*}\Psi >}\right. \\
&&\left. -{\bf \hat{y}}^{\prime \prime }\frac{\left\langle \left[ \left( \Pi
_{y}^{\prime \prime 2}-\Pi _{x}^{\prime \prime 2}\right) d\right] ^{*}\left(
\Pi _{y}^{\prime \prime }d\right) +\left[ \Pi _{y}^{\prime \prime }\left(
\Pi _{y}^{\prime \prime 2}-\Pi _{x}^{\prime \prime 2}\right) d\right]
^{*}d+c.c.\right\rangle }{<\Psi ^{*}\Psi >}\right\}
\end{eqnarray*}
Summing up the corrections, one obtains the corrected current as : 
\[
J=J^{a}+J^{b}={\bf j}+\eta ^{\prime }\delta {\bf j}={\bf j}+\eta ^{\prime
}\left( \delta {\bf j}_{1}+\delta {\bf j}_{2}+\delta {\bf j}_{3}\right) 
\]
Here we have used the fact that the operator $\Pi $ is Hermitian, and the
curly bracket denotes anticommutator. Note that although the total wave
function is a linear combination of $\Psi _{n}(x,y),$ after averaging over
the 2D space all the components decouple due to the $\exp (ik_{n}y)$ factors
and the fact that the current is quadratic in $\Psi _{n}$. This means that
it suffices to consider only one of the components to calculate ${\bf J}$.
The results for the three contributions to the current correction are:

\begin{equation}
\frac{-2%
\mathop{\rm Re}%
<\Psi ^{*}\delta \Psi >}{<\Psi ^{*}\Psi >}{\bf J}=\frac{e^{*}\alpha _{d}}{%
\beta }\frac{1}{\beta _{A}^{0}}\left\{ -g^{4}\left[ 1+\frac{2}{3}\cos
4\Theta \right] +4g^{2}\right\} \left( \gamma {\bf v}\times {\bf \hat{z}}%
\right)
\end{equation}
\begin{equation}
\frac{e^{*}\alpha _{d}}{4\beta m_{d}}\frac{1}{\beta _{A}^{0}}\frac{4%
\mathop{\rm Re}%
<\delta \Psi ^{*}\Pi \Psi >}{<\Psi ^{*}\Psi >}=\frac{e^{*}\alpha _{d}}{\beta 
}\frac{1}{\beta _{A}^{0}}\left\{ g^{4}\left[ 1+\frac{2}{3}\cos 4\Theta
\right] -4g^{2}-2\right\} \left( \gamma {\bf v}\times {\bf \hat{z}}\right)
\end{equation}
\begin{eqnarray}
&&\frac{e^{*}\alpha _{d}}{2\beta }\frac{l_{H}^{2}}{m_{d}}\frac{1}{\beta
_{A}^{0}}\left\{ {\bf \hat{x}}^{\prime \prime }\frac{2%
\mathop{\rm Re}%
\left\langle \Psi ^{*}\left\{ \Pi _{y}^{\prime \prime 2}-\Pi _{x}^{\prime
\prime 2},\Pi _{x}^{\prime \prime }\right\} \Psi \right\rangle }{<\Psi
^{*}\Psi >}-{\bf \hat{y}}^{\prime \prime }\frac{2%
\mathop{\rm Re}%
\left\langle \Psi ^{*}\left\{ \Pi _{y}^{\prime \prime 2}-\Pi _{x}^{\prime
\prime 2},\Pi _{y}^{\prime \prime }\right\} \Psi \right\rangle }{<\Psi
^{*}\Psi >}\right\}  \nonumber \\
&=&\frac{e^{*}\alpha _{d}}{\beta }\frac{1}{\beta _{A}^{0}}\left\{ \left[
g^{2}\left( 1+\cos 4\Theta \right) +2\right] \left( \gamma {\bf v}\times 
{\bf \hat{z}}\right) -g^{2}\sin 4\Theta \left( \gamma {\bf v}\right) \right\}
\end{eqnarray}
where $\Theta \equiv \theta -\varphi $ as before. Summing them up we got our
final expression:

\begin{eqnarray*}
\delta {\bf j} &=&{\bf -}\frac{\beta _{A}^{1}}{\beta _{A}^{0}}{\bf j}+\frac{%
e^{*}\alpha _{d}}{\beta }\frac{1}{\beta _{A}^{0}}g^{2}\left( 1+\cos 4\Theta
\right) \left( \gamma {\bf v}\times {\bf \hat{z}}\right) \\
&&-\frac{e^{*}\alpha _{d}}{\beta }\frac{1}{\beta _{A}^{0}}g^{2}\sin 4\Theta
\left( \gamma {\bf v}\right)
\end{eqnarray*}
From this, one obtains the simple results Eq.(\ref{Jdir},\ref{JHall})
advertised earlier. Note that all the $g^{4}$ terms are cancelled.

\section{Concluding remarks}

Instead of summarizing the results (which has been done in Section I), we
briefly comment on the possibility of observation of various phenomena
quantitatively discussed in this paper.

\subparagraph{\protect\smallskip Internal structure of a single anisotropic
vortex}

Although direct observation of the order parameter using scanning -
tunneling-microscopy (STM)\cite{Hess}, or the magnetic field distribution 
\cite{Blatter} using electron holography \cite{Tonomura2} or other
techniques is possible, the detailed effects hotly debated by theoreticians
(where the zeroes of the s field are located, small distance asymptotics)
probably do not have a significant impact on such experiments. One also
should note that the Ginzburg - Landau framework adopted here might not be
applicable close to the vortex center where microscopic excitation spectrum
becomes important. An approach using elements of the microscopic theory (via
Bogoliubov - DeGennes equations along the lines of \cite{Gygi} and\cite
{Ichioka}) will be necessary.

\subparagraph{Structure of static and moving anisotropic vortex lattice}

Static vortex lattice has been observed using small angle neutron scattering 
\cite{Keimer} and tunneling spectroscopy \cite{Maggio}.\smallskip Although
moving vortices have been directly observed using electron tomography \cite
{Tonomura}, to our knowledge the shape and orientation of moving large
bundles hasn't been observed as yet. Moving vortex lattice is much more
sensitive to pinning effects than the static lattice. In the static case
pinning can just slightly distort or cause breakup of the crystal to smaller
peaces. For the moving flux lattice pinning is expected to be much more
significant. The orientation effect that we predict is very small, but the
asymmetry in magnetization might be quite significant.

We found the transtion point for the parameter $\eta ^{\prime }=\eta
m_{d}e^{*}H$ is at $\eta _{c}^{\prime }=0.0235$. This transition between the
rectangular and the square lattices might be seen in neutron scattering
experiments, since the square lattice has higher symmetry (number of spots
is reduced to four at the transition). Note that by increasing magnetic
field the critical $\eta _{c}^{\prime }$ can be exceeded without changing
the sample ($\eta $ is independent of magnetic field). 

\subparagraph{Static transition to the normal phase}

As is well known, in the presence of fluctuations, the second order phase
transition from superconducting to normal state becomes a weakly first order
melting line into the vortex liquid. This is the reason that the present
study of the diagram will be useless for BSCCO which has a relatively large
Ginzburg number. For YBCO and the low temperature superconductors, the
curvature of the phase transition line can in principle provide an estimate
of $\eta $ with reservations mentioned in the end of Subsection IV.C. 

\subparagraph{\protect\vspace{1pt}Dynamic phase diagram}

Dynamical phase diagram, namely transition from the moving lattice to normal
(or moving liquid) should be complicated by pinning effects. However,
provided these could be overcome a number of interesting effects could be
observed. First, even neglecting the rotational symmetry breaking effects,
there are a number of peculiarities associated with normal - superconducting
boundary noted by Thompson and Hu like the existence of a metastable normal
state and the unstable superconductive state. As far as we know, these
peculiarities haven't been convincingly observed \cite{Huremark} in low $%
T_{c}$ materials. It would be interesting to reconsider this question for
the high $T_{c}$ materials.

In the s - wave case however, the phase diagram cannot depend on orientation
of the current. We calculated this orientation dependance on the angle
between the atomic lattice and the direction of current or electric field to
first order in $\eta $, see Eq.(\ref{line},\ref{line1},\ref{line2}). New
effects include the change in slope of $H_{c2}$ as function of temperature,
not only in curvature.

\paragraph{Nonlinear I-V curves and magnetization}

One should be able to measure currents in the same sample oriented
differently with respect to the atomic crystal. Note that the effect can be
seen in low temperature anisotropic superconductors, not necessarily in
YBCO. The simplicity of the expressions for both direct and Hall current
Eqs.(\ref{Jdir},\ref{JHall}) calls for some special ways to verify it
experimentally. The angular dependence of the magnetization near the
transition (given by Eqs.(\ref{magn},\ref{beta1})) might be large enough to
be measurable.

\smallskip There are number of limitations of our approach which can be
lifted by possible extensions. One of them is the assumption of exact
fourfold symmetry. Deviations from it in a form of different coefficients of
the gradient terms in x and y directions have already been studied recently 
\cite{Ting4} using the two field formalism. If they happen to be small, they
can be easily added perturbatively. These effects of explicit breaking are
clearly quite different from those of spontaneous breaking of the fourfold
symmetry studied here. Our results for the latices are limited to fields
close to $H_{c2}$ only. It is possible, although more difficult to extend
them to lower magnetic fields. Another interesting direction is the
influence of the anisotropy on vortex fluctuations in the lattice. We hope
to address these issues in the future.

Also anisotropy influences fluctuations not considered here. In addition,
the effective one component approach allows to consider possibilities not
apparent within the two field one. For example, the coefficient $\eta ,$ in
principle, can be negative despite the fact that within the two field
formalism it should be positive. Twinning is expected to reduce the value of
the parameter.

\appendix
\section*{Comparison of the two field and the one field
results for single vortex}

In the two field formulation, the small $r$ asymptotics of the solution for
the d - wave component of the isolated vortex is given by :

\begin{equation}
d(r,\phi )\simeq \left( d_{1}r+d_{3}r^{3}\right) e^{i\phi }
\label{dtwofield}
\end{equation}
where the subleading term coefficients is

\[
d_{3}=-\frac{d_{1}}{8\xi _{d}^{2}}\left[ 1+\frac{h_{0}}{H_{c2}(0)}\right]
\simeq -\frac{d_{1}}{8\xi _{d}^{2}}, 
\]
neglecting terms proportional to $h_{0}/H_{c2}(0)$ \cite{Berlinsky2}. The $s$%
-component asymptotics is: 
\begin{eqnarray}
s(r,\phi ) &\simeq &-{{\frac{\gamma _{v}}{\alpha _{s}}}}(\Pi _{y}^{2}-\Pi
_{x}^{2})d(r)  \nonumber \\
&\simeq &-{{\frac{\gamma _{v}}{\alpha _{s}}}}\left( 4d_{3}re^{-i\phi
}\right) =+\frac{1}{2}\left( \frac{\gamma _{v}}{\alpha _{s}\xi _{d}^{2}}%
\right) d_{1}re^{-i\phi }  \label{stwofield}
\end{eqnarray}

The expression Eq.(\ref{stwofield}) is different from what was obtained in 
\cite{Berlinsky2}, Eq. (22), but qualitatively the behavior is not affected.
We also found that it doesn't follow from their Eq. (19), because to the
same order of approximation \cite{Berlinsky2} had a non-vanishing term
proportional to $e^{3i\phi }$. Nevertheless the concluding statement in \cite
{Berlinsky2} is basically correct. Following the same argument leading to an
estimate of the maximal amplitude of $s(r)$ as in \cite{Berlinsky2} we
obtained 
\[
\frac{s_{\max }}{d_{0}}\simeq \frac{1}{4}\left( \frac{\gamma _{v}}{\alpha
_{s}\xi _{d}^{2}}\right) 
\]
apparently this correction accounts for the 20\% error cited in \cite
{Berlinsky2}.

The asymptotic form of the wave function was used to make a topological
argument about poles in the s - wave. Due to the different winding number of
small $r$ and large $r$ asymptotics of $s(r,\phi )$, there must exist four
poles in the intermediate region. This was shown numerically in \cite
{Berlinsky2}. Ting {\it et al }\cite{Ting2}{\it \ }however performed a
similar calculation, but didn't get the poles. Our calculation, which is
much simpler than the two field one, confirms the former and shows clearly
four poles on the $x$ and $y$ axis, independent of what kind of approximate $%
d-$component wave function one chooses. We suspect that the numerical
simulation in \cite{Ting2} was not sensitive enough to resolve these poles 
\cite{Tingcomment}.

\acknowledgments
Authors are very grateful to our colleagues here in Hsinchu C.C. Chi, Y.S.
Guo, M.K. Wu, Vincent Yang, S.Y. Hsu for encouragement and discussions and
to Profs. C.R. Hu, C.S. Ting and F.C. Zhang for discussions and
correspondence. The work of B. R. supported by NSC grant 86-2112-M009-034T,
of D.C. and C.L. Wu 86-2112-M009-034T and of C. -Y. M. 86-2112-M009-034T.

\newpage
\noindent
{\bf\large Figure Captions}
\newline

\noindent
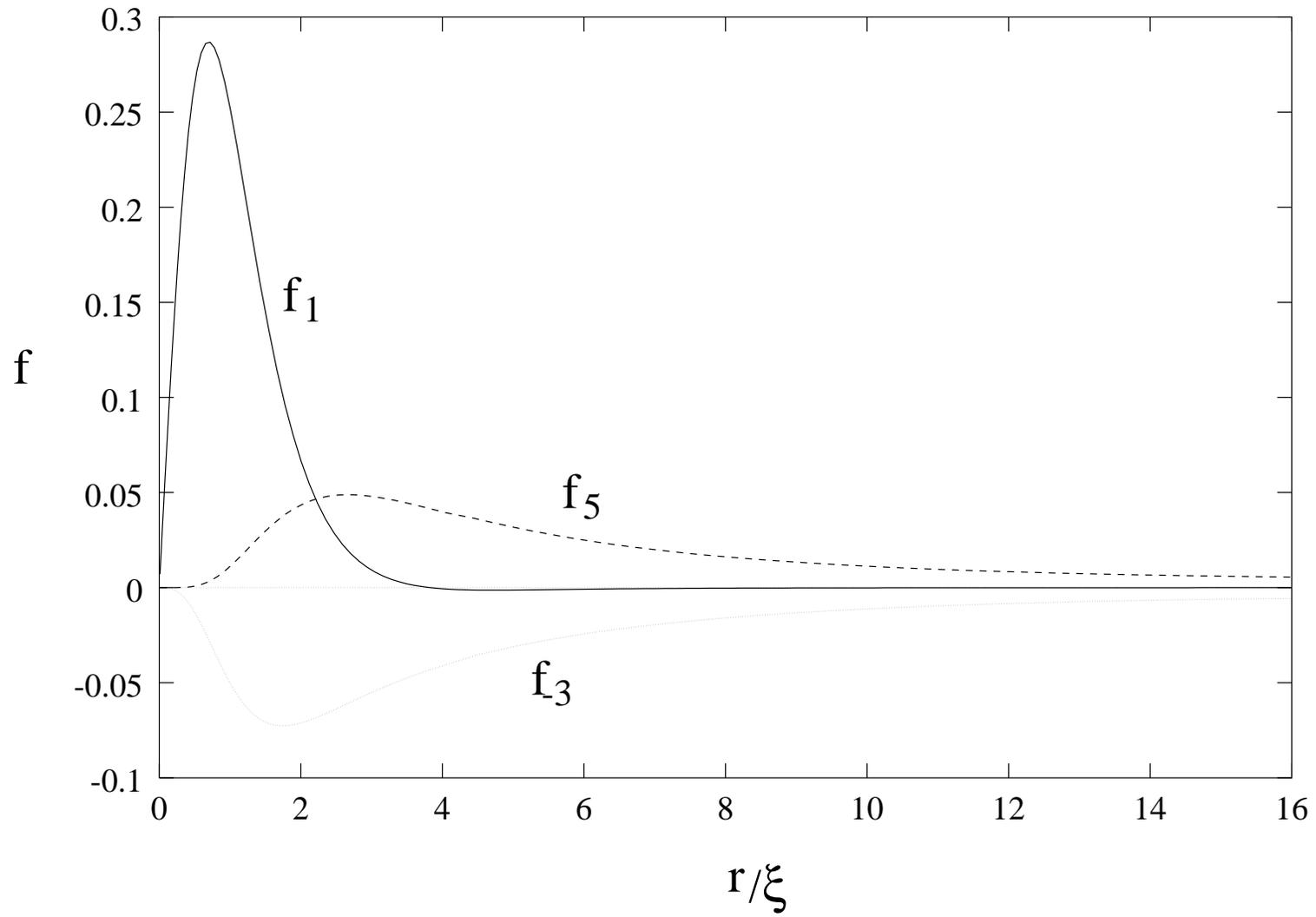
{\sc FIG. 1} A single vortex solution of the one component GL equation. The
coefficient function $f_{1},f_{-3},f_{5}$ for the harmonics $e^{i\phi
},e^{-3i\phi },e^{5i\phi }$, respectively. $f_{i}$'s are given in units of $%
\Psi _{0}=\alpha _{d}/2\beta $, and $r$ is given in units of $\xi _{d}$. See
Eq. (\ref{f3}),(\ref{f5}),(\ref{f1})
\newline

\noindent
{\sc FIG. 2} The {\it d}-field of a single vortex for $\eta =0.15$. Only the
absolute value of the {\it d} field in units of $\Psi _{0}$ is shown. (a)
Contour plot. (b) Three dimensional plot.
\newline

\noindent
{\sc FIG. 3} The {\it s}-field of a single vortex for $\eta =0.15$. (a) Contour
plot. (b) Three dimensional plot. Note that there are four singularities on
which the {\it s}-field vanishes.
\newline

\noindent
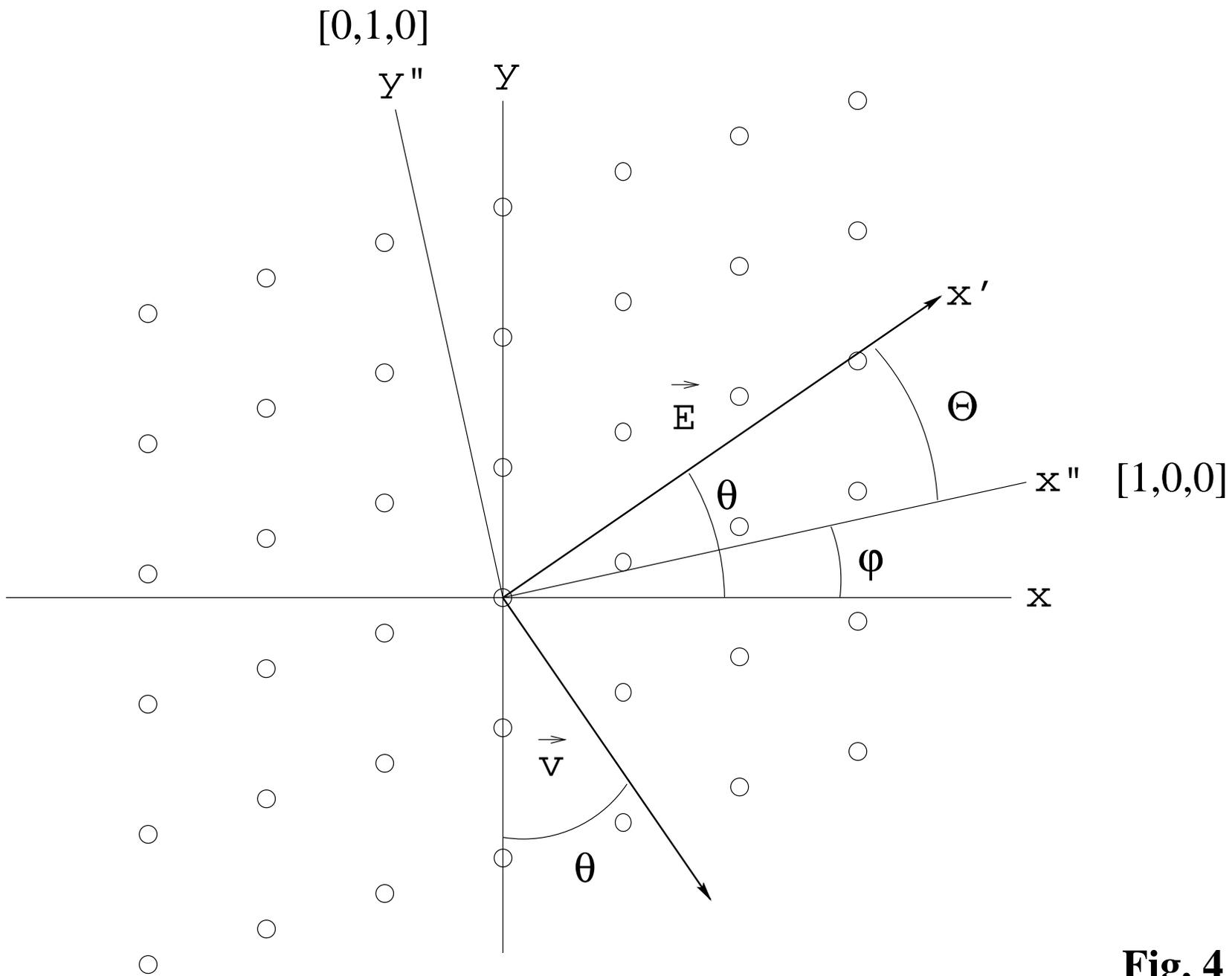
{\sc Fig. 4} The coordinate system used in our calculations, this defines the angles $\theta ,\varphi $ and $\Theta$.
\newline

\noindent
{\sc FIG. 5} The Abrikosov parameter $\beta _{A}$ as a function of the lattice
parameters $(\rho ,\sigma )$ (defined in Eq.(\ref{zeta})). There are three
degenerate local minima. Oblique lattices are on the lines $\rho =1/2$ and $%
\rho ^{2}+\sigma ^{2}=1$. The two points A and B are related by $\rho
\rightarrow 1/\rho $ and therefore represents the same lattice. Point C
represents the same rectangular lattice rotated by $90^{o}$.
\newline

\noindent
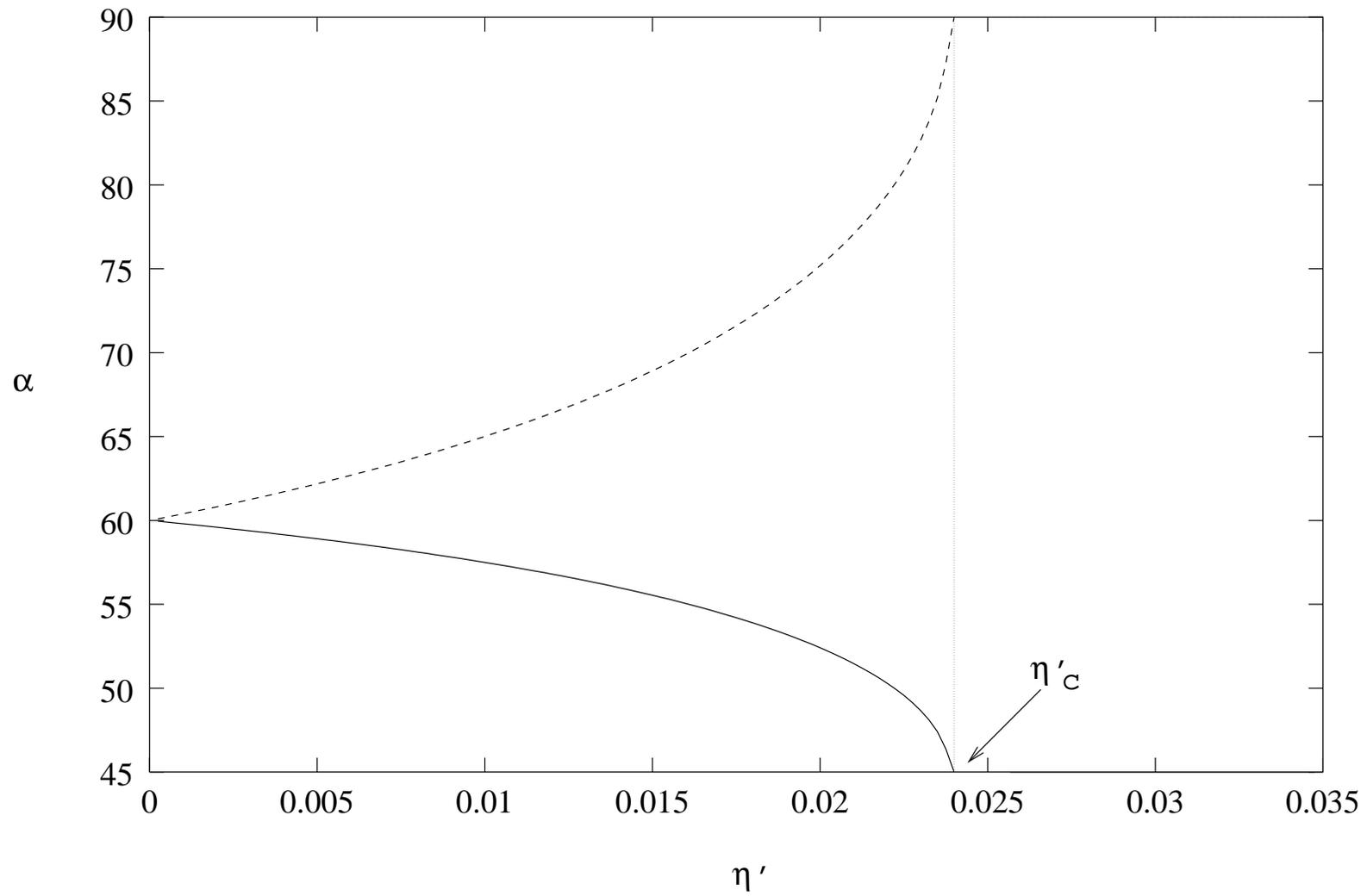
{\sc FIG. 6} The angle $\alpha $ as a function of $\eta ^{\prime }$, the two
branches correspond to lattices related by a rotation of $90^{o}$. A
continuous transition from the rectangular lattice to the square lattice happens at $%
\eta _{c}^{\prime }=0.235$.
\newline

\noindent
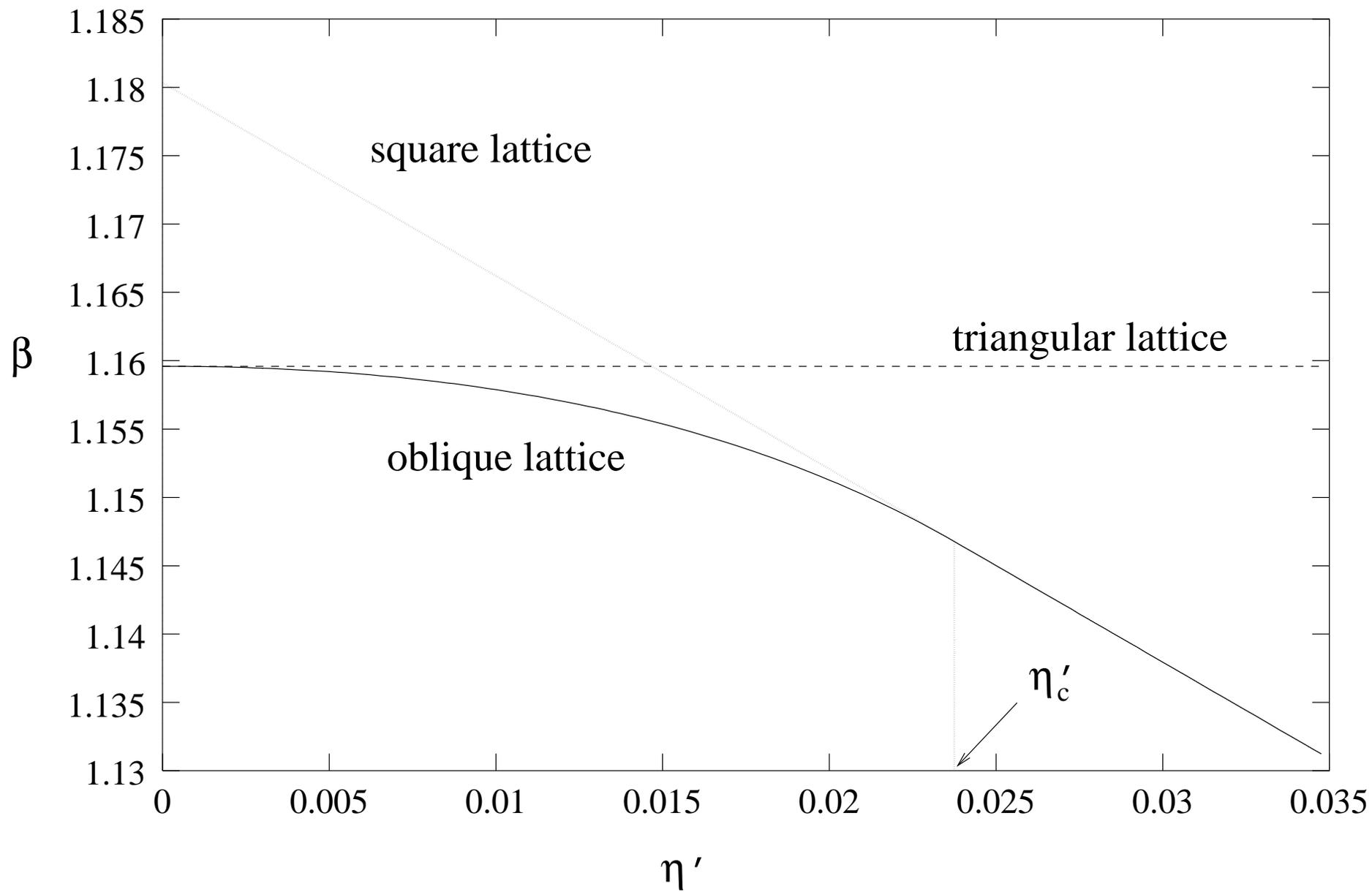
{\sc FIG. 7} The Abrikosov parameter $\beta _{A}$ as a function of $\eta ^{\prime }$
for triangular, square and optimal rectangular body centered lattices,respectively. At the transition point 
$\eta _{c}^{\prime }$, the rectangular lattice is taken over by the square lattice. 
Note that $\eta^{\prime}$ is proportional to the magenetic field $H$.

\end{document}